\documentclass[conference]{IEEEtran}

\ifCLASSOPTIONcompsoc
  \usepackage[nocompress]{cite}
\else
  \usepackage{cite}
\fi

\usepackage{amssymb}
\RequirePackage{shellesc} 
\PassOptionsToPackage{colorlinks=true,
    linkcolor=blue,
    urlcolor=blue,
    citecolor=blue,
    anchorcolor=blue}{hyperref}

\usepackage{amsmath, amssymb}
\usepackage{booktabs}
\usepackage{caption}
\usepackage{subcaption}
\usepackage{colortbl}
\usepackage{multirow}
\usepackage{tikz}
\usepackage{enumitem}
\usepackage{siunitx}
\usepackage[linesnumbered,ruled,vlined]{algorithm2e}
\usepackage[all]{nowidow}
\usepackage{diagbox, eqparbox, hhline}
\usepackage{fontawesome}
\usepackage{xcolor} 
\usepackage[numbers,sort]{natbib}
\usepackage{tcolorbox}
\usepackage{hyperref}

\SetCommentSty{mycommfont}

\definecolor{Gray}{gray}{0.9}

\newcommand{\eg}{e.g., }
\newcommand{\ie}{i.e., }

\newcommand{\shortsection}[2][.]{\vspace{1mm}\noindent\textbf{#2#1}}

\newlist{rqlist}{enumerate}{3}
\setlist[rqlist]{label=\textbf{RQ\arabic*}.,before=\raggedright,leftmargin=40pt,ref=RQ\arabic*}

\newcounter{takeaway}
\newcommand{\takeaway}{\refstepcounter{takeaway}\textbf{Takeaway~\arabic{takeaway}}}

\RequirePackage[normalem]{ulem}
\RequirePackage{color}
\definecolor{RED}{rgb}{1,0,0}
\definecolor{BLUE}{rgb}{0,0,1}

\newif{\ifanonymous}

\newif{\ifrevision}

\begin{document}

\newcommand{\itembase}[1]{\setlength{\itemsep}{#1}}

\date{}

\title{\Large \bf Targeting Alignment: Extracting Safety Classifiers of Aligned LLMs}
\ifanonymous
\author{\rm Anonymous Authors}
\else
\author{
\IEEEauthorblockN{
Jean-Charles Noirot Ferrand\IEEEauthorrefmark{1},
Yohan Beugin\IEEEauthorrefmark{1},
Eric Pauley\IEEEauthorrefmark{2},
Ryan Sheatsley\IEEEauthorrefmark{1},
Patrick McDaniel\IEEEauthorrefmark{1}
}
\IEEEauthorblockA{\IEEEauthorrefmark{1}
Department of Computer Sciences\\
University of Wisconsin--Madison\\
Madison, WI, USA\\
\{jcnf,ybeugin,sheatsley,mcdaniel\}@cs.wisc.edu
}
\IEEEauthorblockA{\IEEEauthorrefmark{2}
Department of Computer Science\\
Virginia Tech\\
Blacksburg, VA, USA\\
pauley@cs.vt.edu
}
}

\fi
\maketitle

\begin{abstract}
Alignment in large language models (LLMs) is used to enforce guidelines such as safety.
Yet, alignment fails in the face of \textit{jailbreak} attacks that modify inputs to induce unsafe outputs.
In this paper, we introduce and evaluate a new technique for jailbreak attacks. We observe that alignment embeds a safety classifier in the LLM responsible for deciding between refusal and compliance, and seek to extract an approximation of this classifier: a surrogate classifier. To this end, we build \textit{candidate} classifiers from subsets of the LLM.
We first evaluate the degree to which candidate classifiers approximate the LLM's safety classifier in benign and adversarial settings. Then, we attack the candidates and measure how well the resulting adversarial inputs transfer to the LLM.   
Our evaluation shows that the best candidates achieve accurate agreement (an $F_1$ score above 80\%) using as little as 20\% of the model architecture. Further, we find that attacks mounted on the surrogate classifiers can be transferred to the LLM with high success. For example, a surrogate using only 50\% of the Llama 2 model achieved an attack success rate (ASR) of 70\% with half the memory footprint and runtime---a substantial improvement over attacking the LLM directly, where we only observed a 22\% ASR.
These results show that extracting surrogate classifiers is an effective and efficient means for modeling (and therein addressing) the vulnerability of aligned models to jailbreaking attacks. The code is available at \url{https://github.com/jcnf0/targeting-alignment}.
\end{abstract}

\section{Introduction}\label{section:introduction}
Improvements in machine learning---specifically the introduction of the transformer architectures~\cite{vaswaniAttentionAllYou2023}---have led to increasingly powerful large language models (LLMs).  Whether they are accessed through an API or interface (\eg GPT~\cite{openai2024gpt4technicalreport}, Claude~\cite{anthropicIntroducingClaude}) or published as open-source (\eg Llama 2~\cite{touvronLlama2Open2023}, Llama 3~\cite{dubeyLlama3Herd2024}, Qwen~\cite{teamQwen25PartyFoundation2024}, etc.), these models have become the de facto tool for tasks involving natural language. They serve as foundations for new tools, in which they are augmented with new capabilities~\cite{anthropicIntroducingComputerUse, openaiIntroducingChatGPTSearch} or fine-tuned on a downstream task~\cite{huLLMAdaptersAdapterFamily2023}. 
In adapting these models to tasks, LLMs undergo an alignment process in which they are further trained (refined) to satisfy safety objectives~\cite{touvronLlama2Open2023, rafailovDirectPreferenceOptimization2023}. For example, alignment guidelines often prevent a model from generating responses that would be offensive, discriminatory, or harmful (\ie unsafe). It is acknowledged that alignment can (and often does) fail in adversarial settings, resulting in a \textit{jailbreak}: an unsafe input being accepted as safe. Indeed, many studies have explored the automated generation of jailbreak inputs, \ie adversarial examples~\cite{zouUniversalTransferableAdversarial2023,liuAutoDANGeneratingStealthy2023,liaoAmpleGCGLearningUniversal2024}. However, these attacks, specifically in white-box settings, suffer from two major drawbacks: they are inefficient (\ie high runtime and memory footprint) and employ various heuristics (\eg maximize the likelihood of a specific output sentence~\cite{zouUniversalTransferableAdversarial2023}, apply fixed changes to the input~\cite{yongLowResourceLanguagesJailbreak2024,jiangArtPromptASCIIArtbased2024}, etc.) that limit the search space of adversarial inputs and thus lower their efficacy and their ability to assess the robustness of models.

\begin{figure}[t]
    \centering
    \includegraphics[width=\columnwidth]{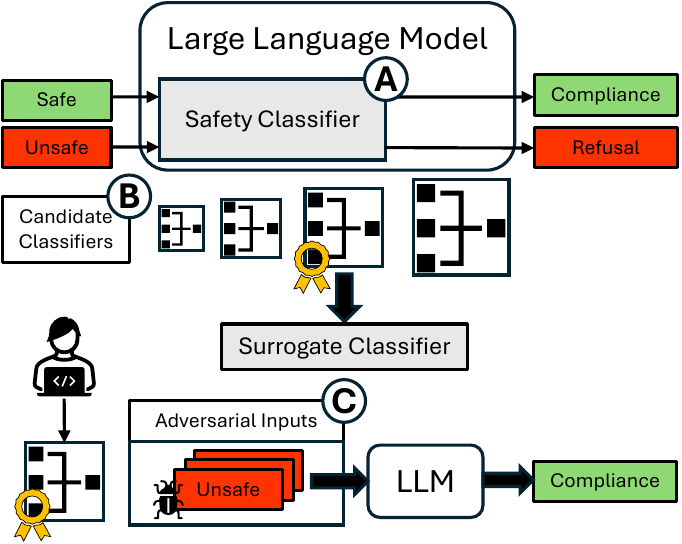}
    \caption{In this work, we (A) hypothesize that alignment embeds a safety classifier in LLMs responsible for the classification of safe and unsafe inputs. Then, we design an approach that (B) builds candidate classifiers  using the model structure. We evaluate these candidates in benign and adversarial settings to select the best, called the surrogate classifier. Finally, we (C) attack the surrogate classifier to generate adversarial inputs that transfer to the LLM.}
    \label{fig:teaser-figure}
\end{figure}

In this paper, we posit a novel approach to reason about LLM alignment in the context of jailbreak attacks (see ~\autoref{fig:teaser-figure}). Here, we hypothesize that alignment embeds a ``safety'' classifier into an LLM that determines whether an input prompt is safe (compliant with alignment goals) or unsafe (refusal with alignment). In order to evaluate this hypothesis, we extract an approximation of the model's classifier (called a {\it surrogate classifier}) and evaluate its utility in assessing a model's robustness to adversarial inputs. We explore how a surrogate classifier can be extracted, and thereafter how the surrogate classifier can be used to assess robustness. In this, we demonstrate that surrogate classifiers are an efficient and accurate means of evaluating the robustness of a target LLM.

We begin by exploring methods for extracting surrogate classifiers. Here, our approach systematically identifies candidates from subsets of the LLM's architecture. Candidates are built by (a) finding a structure, \ie a subset of the architecture of the model, that best represents the embedded safety classifier, (b) adding a classification head, and (c) training this classification head on unsafe and safe inputs to map the features extracted from the structure to the model predictions (positive for refusal, negative for compliance). The candidates are then evaluated in benign and adversarial settings. For the benign settings, we measure how well the candidate classifier agrees with the model's safety classifier on safe and unsafe inputs. For the adversarial settings, we measure transferability of adversarial inputs between the LLM and candidate classifiers to determine how well the latter capture the decision of the embedded safety classifier.

We perform a detailed analysis of the efficacy of our approach over 4 models and 2 datasets. We first evaluate candidates with respect to benign inputs, assessing whether each candidate classifier agrees with the safety classification of the LLM. We find that the best candidates achieve accurate agreement (an \(F_1\) score above 80\%) using as little as 20\% of the model architecture. Next, we evaluate the candidate classifiers with respect to transferability of adversarial inputs (\ie as generated using best-in-class jailbreaking algorithms) from the candidates onto the LLM. These latter experiments show that some candidate classifiers are extremely effective and efficient tools for targeting LLMs. For example, a surrogate classifier using only 50\% of the Llama 2 model achieved an attack success rate (ASR) of 70\%---a substantial improvement over attacking the LLM directly, where we observed a 22\% ASR---with less than half the runtime (350s/sample against 850s/sample) and the memory footprint (14~GB of VRAM usage against 31~GB).

Our results suggest that investigating the robustness can be furthered (and in many cases made more efficient) by extracting surrogate classifiers that target alignment tasks. In this paper, we have found that approximations of LLMs may be sufficient to explore the security of their deployment, and at the same time offer potential vectors for developing more effective attacks on alignment. 

Our contributions are as follows:
\begin{itemize}
    \item We introduce a novel paradigm for jailbreak attacks that extracts a surrogate classifier from a subset of the LLM's weights and attacks it.

    \item We evaluate this paradigm on several state-of-the-art LLMs, showing that attacking the surrogate classifier and transferring to the LLM is more successful than attacking the LLM directly.

    \item We show that studying alignment robustness is scalable, as our approach significantly reduces the memory footprint and runtime of the attack.
    
\end{itemize}

\section{Background}\label{section:background}
\subsection{Large Language Models (LLMs)}\label{section:llm}
\shortsection{Large language model} An LLM models the conditional probability of the next token (a unit of text for the model) with respect to previous tokens. Let \(V\) be the vocabulary of the tokens and \(V^\ast =\bigsqcup\limits_{n=1}^{N}V^n\) the corresponding input space of sequences, where \(N\) denotes the context window of the model (maximum amount of tokens that can be processed by the model). Given a sequence of tokens \(x = x_1 x_2 \dots x_T \in V^{\ast}\), the model aims to produce the next token \(x_{T+1}\) by approximating the probability distribution.
\begin{equation}
    p(x_{T+1}|x_1 x_2 \dots x_T)
\end{equation}
LLMs learn a vector representation of words called the embedding space. Given a sequence of tokens \(x=x_1 x_2 \dots x_T\), each token is assigned to an embedding, resulting in a sequence of vectors \(\{h_t\}_{t\in\{1,\dots,T\}}\) all in \(\mathbb{R}^d\), where \(d\) is the dimension of the embedding space. This study focuses on aligned chat models, which are almost exclusively of the same architecture: a sequence of \textit{decoders}. The \(i\)-th decoder transforms a sequence of embeddings \(h^{(i)}\) into another \(h^{(i+1)}\), keeping the same sequence length. If the model is made of \(D\) decoders, the probability distribution on the next token is obtained by applying a linear layer followed by a softmax function on the last output embedding of the last decoder, \(h_{T}^{(D)}\), \ie \(p(\cdot|x_1 x_2 \dots x_T) = \text{softmax}(Ah_{T}^{(D)})\), where \(A\in\mathbb{R}^{d\times|V|}\) is the learned matrix that maps the embeddings to the scores (or logits) for each token.


\subsection{Alignment and Safety}\label{section:alignment}
\shortsection{Alignment} Since the training data used for LLMs comes from diverse sources, biases and undesirable behaviors emerge~\cite{gallegosBiasFairnessLarge2024}, leading to misaligned pretrained models~\cite{anwar2024foundational}. In order to prevent those behaviors from happening, models often go through an alignment process that regulates their outputs according to given guidelines. Several methods exist and are not mutually exclusive~\cite{touvronLlama2Open2023}. Approaches such as supervised fine-tuning (SFT) use human-generated sample responses, while reinforcement learning with human feedback (RLHF)~\cite{bai2022traininghelpfulharmlessassistant,christiano2017deepreinforcementlearninghuman} trains a neural network that acts as a reward for the LLM. Since training another reward model may be costly, other approaches such as direct preference optimization (DPO)~\cite{rafailovDirectPreferenceOptimization2023} uses the LLM as its own reward model.

\shortsection{Guidelines and taxonomies} The existence of alignment as a technique is a direct product of the complexity of identifying what is precisely expected from models (\ie their objective). Safety is a broad term; therefore, previous work has identified taxonomies~\cite{inanLlamaGuardLLMbased2023} that represent a wide range of unsafe behaviors. This allows researchers to evaluate alignment on a finer scale and understand where it could be improved. For example, when the prompts are decomposed into different categories such as self-harm, privacy, or harassment, models tend to refuse prompts related to self-harm and accept prompts related to privacy~\cite{cuiORBenchOverRefusalBenchmark2024}.

\shortsection{Identifying unsafe outputs} Classifying LLM outputs is challenging. Since earlier aligned models created strong refusal responses, a naive solution is to test whether certain refusal keywords are in the output (\eg \textit{"Sorry"}, \textit{"as a responsible AI"}, etc.)~\cite{zouUniversalTransferableAdversarial2023}. However, this method is less accurate for more recent models that were aligned with a different approach than refusal training~\cite{yuan2025hardrefusalssafecompletionsoutputcentric}. Alternatively, LLM judge models are trained to classify the safety of outputs with more~\cite{inanLlamaGuardLLMbased2023} or less~\cite{mazeikaHarmBenchStandardizedEvaluation2024} granularity on the classes they predict, similarly to the taxonomies explained above. A limitation of judge models is that they can also be attacked~\cite{mangaokarPRPPropagatingUniversal2024,rainaLLMasaJudgeRobustInvestigating2024}, leading to a possible high rate of false negatives (unsafe outputs judged as safe).

\subsection{Attacks on LLMs}
LLMs, like any machine learning model, are prone to attacks. While several exploits exist (\eg model-stealing~\cite{carliniStealingPartProduction2024}, text misclassification, etc.~\cite{zhuPromptRobustEvaluatingRobustness2024}), the \textit{jailbreak} exploit remains the most prevalent. Jailbreak refers to the compliance of an aligned LLM for an unsafe query, as defined by the alignment guidelines. Jailbreak attacks fall into one of two threat models: the white-box threat model, in which model parameters are known, and the black-box threat model, in which only access to the output of the model is assumed. Beyond their use for malicious behavior, jailbreak attacks are also used to test LLMs before their deployment through red-teaming~\cite{verma2025operationalizing}.

\shortsection{White-box} These attacks have been studied on open-weights models such as Llama 2~\cite{touvronLlama2Open2023}, Llama 3~\cite{dubeyLlama3Herd2024}, or Vicuna~\cite{chiangVicunaOpenSourceChatbot2023}. One of the first white-box attacks, GCG~\cite{zouUniversalTransferableAdversarial2023}, uses greedy coordinate gradient (GCG) to mitigate the absence of mapping from embeddings (\ie vectors) to tokens. Following this, many new white-box algorithms have been introduced~\cite{xuComprehensiveStudyJailbreak2024,chuComprehensiveAssessmentJailbreak2024}, improving existing algorithms~\cite{zhaoAcceleratingGreedyCoordinate2024,paulusAdvPrompterFastAdaptive2024,liaoAmpleGCGLearningUniversal2024,linUnderstandingJailbreakAttacks2024}, using different objectives~\cite{zhouDontSayNo2024}, or using specific properties of the model (\eg logits~\cite{liLockpickingLLMsLogitBased2024}, special tokens~\cite{zhouVirtualContextEnhancing2024}, or generation~\cite{huangCatastrophicJailbreakOpensource2023}). Attacks like GCG have an average attack success rate (ASR) above 50\%~\cite{mazeikaHarmBenchStandardizedEvaluation2024} on most open-source models.

\shortsection{Black-box} These attacks have been widely studied, given their prevalence in practical scenarios~\cite{openai2024gpt4technicalreport,anthropicIntroducingClaude}. These attacks generally employ heuristics (transformations to the initial prompt such as ASCII-art~\cite{jiangArtPromptASCIIArtbased2024}, translation into underrepresented languages~\cite{yongLowResourceLanguagesJailbreak2024}, or using pre-existing templates~\cite{yuDontListenMe}), search algorithms~\cite{chaoJailbreakingBlackBox2024,mehrotraTreeAttacksJailbreaking2024}, or a combination thereof. Black-box attacks achieve remarkable efficacy. For instance, ArtPrompt~\cite{jiangArtPromptASCIIArtbased2024} can achieve ASR as high as 80\% on commercial LLMs such as GPT-3.5.
\section{Methodology}\label{methodology}

This section outlines our approach for extracting and attacking surrogate classifiers of LLMs. Before we detail our methodology, we first explain a preliminary experiment whose results inform our approach.

\subsection{Empirical Evidence of a Safety Classifier}\label{section:subspace-analysis}
As explained previously, LLMs are aligned but are prone to attacks. Regardless of the threat model, jailbreak attacks have given limited understanding of where alignment fails, since they involve heuristics in either the adversarial objective or the perturbation they apply to the input. In this work, we hypothesize that alignment embeds a safety classifier in the model, whose existence we outline next. By extracting this safety classifier, we can create a new class of optimization-based attacks that reduces jailbreaking to misclassification on this surrogate classifier.

\shortsection{Motivation of existence} While refusal from aligned models started as sets of predefined outputs (\eg \textit{``I cannot fulfill your request. I'm just an AI''}), this idea was challenged~\cite{qi2025safety} and recent efforts~\cite{yuan2025hardrefusalssafecompletionsoutputcentric} have tried to pivot away from refusal training and focus on safe-completions. Regardless, given an input, the model has to decide whether an input is safe or unsafe. We assume the following: the last intermediate state before the output is enough information to know whether the output will be a refusal or not. Under this assumption, there exists a set of refusal intermediate states \(H_r\) such that, for a given input sequence of tokens \(x=x_1 x_2 \dots x_T\in V^\ast\), \(h^{(D)}_{T}\in H_r\) if and only if the output answer is a refusal. Thus, classifying unsafe inputs reduces to verifying \(h^{(D)}_{T}\in H_r\). 

\shortsection{Separation metric} For the classifier to exist, there must be separation between the two classes, \ie inputs classified as unsafe (refusal) and inputs classified as safe (compliance). To demonstrate this, we measure the average silhouette score on the embeddings associated to the safe and unsafe inputs. We use the silhouette score to determine how well the two safety classes cluster and to evaluate whether there is a separation, defined as follows for a given data point \(i\):

\begin{equation}
    s(i) = \frac{d_{\text{inter}}(i) - d_{\text{intra}}(i)}{\max(d_{\text{intra}}(i), d_{\text{inter}}(i))}
\end{equation}
where \(d_{\text{intra}}(i)\) denotes the mean intra-cluster distance (average distance to all other points in the same cluster) and 
\(d_{\text{inter}}(i)\) the mean nearest-cluster distance (average distance to points in the nearest cluster).

Silhouette scores range from \(-1\) to \(1\) where scores below 0 indicate that representations overlap, and scores above 0.25 or 0.5 imply a weak or reasonable separation~\cite{ROUSSEEUW198753}.

\begin{figure}[ht!]
    \centering
    \includegraphics[width=\columnwidth]{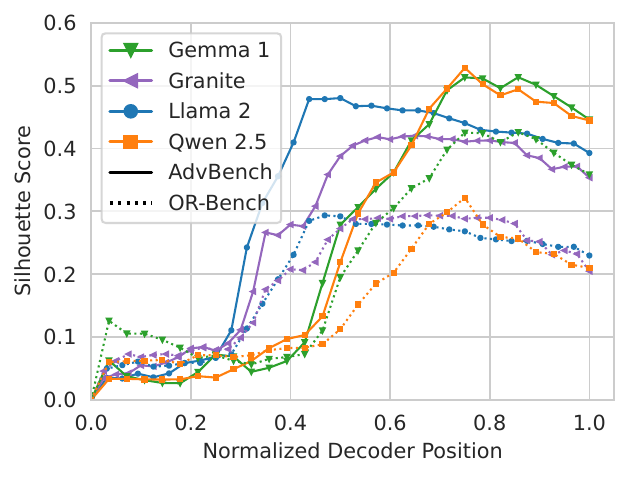}
    \caption{Silhouette score (measure of separation) of unsafe and safe input embeddings for different LLMs.}
    \label{fig:silhouette-full}
\end{figure}

\shortsection{Result} We measure this score on four LLMs and two datasets (see \autoref{section:setup} for the detailed experimental setup). \autoref{fig:silhouette-full} shows the silhouette score of the embeddings for different models and datasets, with respect to the normalized decoder position (ratio between the current decoder position and the total number of decoders).

While it would be expected to see a nondecreasing trend\footnote{Usually, the subsequent layers of classifiers do not lose information on the classification, hence a nondecreasing trend. For LLMs, this classification is only one of many tasks that are performed, which is why we observe a compression of the linear information in later layers.}, we see that after the silhouette score reaches its maximum, it starts to decrease. A possible explanation for this phenomenon is rooted in the fact that models are designed to perform multiple tasks. Therefore, after extracting and integrating the safety information from the input, the other tasks add more information to the embedding that is not related to safety. Thus, the safety separation decreases (\ie the linear information is compressed), but remains high enough for the classification. Besides the difference in magnitude of separation between the two datasets, the trend is the same for all models.

\begin{tcolorbox}[title=\takeaway{}: Existence of a safety classifier]
     Across models and datasets, internal decoders show separation between safe and unsafe inputs that exceeds that of the final decoder, suggesting the existence of an embedded safety classifier that does not span the full model architecture.
\end{tcolorbox}

\subsection{Threat Model}\label{section:threat-model}
Motivated by our preliminary results, our adversary aims to extract an approximation of the safety classifier of a target aligned model, called a surrogate classifier. A successful extraction would allow subsequent attacks to be made with more effective white-box jailbreak approaches, while forgoing the complex computational costs they incur, as we discuss in \autoref{section:implications-wb}. We will extract the surrogate classifier using subsets of the model, thus white-box access is assumed (\ie access to the model weights). In practice, the actor could be a red-teamer that would benefit from the efficiency and efficacy gains of the approach when auditing a model.

In jailbreak attacks, the adversarial objective is to make the model comply with an unsafe input by adding a perturbation. Aligned models overfit on the unsafe inputs. As a result, they can reject safe prompts~\cite{cuiORBenchOverRefusalBenchmark2024}, exhibiting a high sensitivity to certain tokens. While previous work has mostly studied jailbreaks (unsafe inputs identified as compliant), in this work we focus on extracting the surrogate classifier therein enabling attacks that target alignment more closely. To do so, we also study safe inputs identified as unsafe, as we hypothesize that they also provide substantial information on the decision boundary of the classifier. 

\begin{figure*}[ht]
    \centering
    \includegraphics[width=\textwidth,height=4cm]{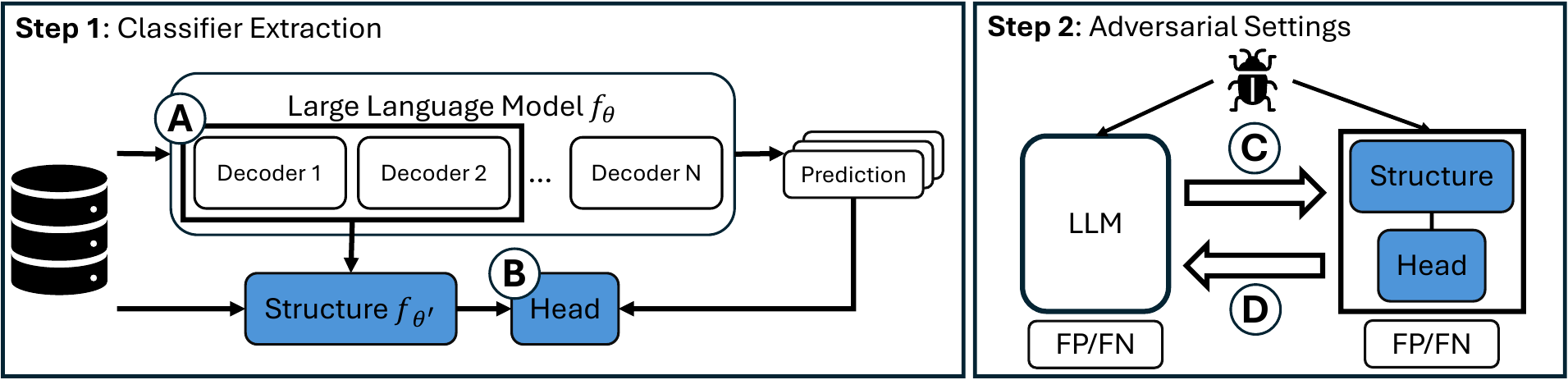}
    \caption{Methodology overview. In the first step, we extract the safety classifier of an LLM by (A) selecting a structure within the model and (B) training a classification head on the predictions of the LLM to create a candidate classifier. We verify its performance in benign settings. Then, we measure the transferability between the LLM and the candidate in both ways: adversarial examples of the LLM to the candidate (C) and adversarial examples of the candidate to the LLM (D).}
    \label{fig:methodology}
\end{figure*}

\subsection{Problem Formulation}\label{section:methodo-problem}
\shortsection{General formulation} Let \(f_\theta\) be an LLM with context window \(N\) and vocabulary \(V\), and let \(\mathcal{R}: x\in V^{\ast}\mapsto C\) be a classification rule that maps the output of the LLM to a class from \(C\). We aim to extract a \textit{classifier} from the LLM through two components: a \textit{structure} and a classification head. A \textit{structure} \(f_{\theta^\ast}, \theta^\ast\subset\theta\) is a subset of the model (\eg a decoder). A classification head \(\mathcal{C}\) maps the representations learned by the structure to a class such that the classifier \(\mathcal{C}\circ f_{\theta^\ast}\) is equivalent to the LLM for the classification task. Formally, for any input \(x\in V^{\ast}\), the following holds:
\begin{equation}\label{eq:perfect-agreement}
    \mathcal{R}(f_\theta (x)) = \mathcal{C}(f_{\theta^\ast}(x)).
\end{equation}

\shortsection{Safety} In this paper, we focus on safety classification, i.e., distinguishing between unsafe and safe inputs. This is a binary classification problem, where unsafe prompts are considered positive observations and safe prompts as negative observations. A jailbreak is then a false negative, while a false positive is ``over-refusal''~\cite{cuiORBenchOverRefusalBenchmark2024}. The classification rule \(\mathcal{R}\) to assign a predicted LLM label is then comparable to how attack success is measured for jailbreak attacks, \ie by verifying refusal or compliance.
The classification rule \(\mathcal{R}\) we chose is a model trained to classify LLM outputs as refusal or compliance as we focus on the type of answer from the LLM, not if it produces unsafe information.

\subsection{Extracting the Classifier}\label{section:methodo-estimating}
In order to extract the classifier, we first need to have a set of candidates. We build a candidate classifier through the following three steps, summarized in \autoref{fig:methodology}:
\begin{enumerate}
    \item Select a subset \(f_{\theta^\ast}\) of the architecture of the LLM \(f_\theta\), which we will refer to as a \textit{structure}.
    \item Collect data points from the structure \((f_{\theta^\ast}(x), \mathcal{R}(f_{\theta}(x)))\) consisting of the features extracted from the structure and the predicted label of the LLM. 
    \item Train a classification head \(\mathcal{C}\) on \((f_{\theta^\ast}(x), \mathcal{R}(f_{\theta}(x)))\).
\end{enumerate}
The result, \(\mathcal{C}\circ f_{\theta^\ast}\), will be termed a \textit{candidate} classifier.

\shortsection{Structures} As explained in \autoref{section:llm}, most LLMs are built as sequences of decoders: processing units that take a sequence of vectors and output another sequence of vectors of the same length. LLMs can be scaled either in depth by adding decoders or in width by increasing the dimensionality of embeddings. Therefore, there are many ways to ``cut'' the model and thus many possible structures. It is clear that there is a separation between unsafe and safe representations at the end of aligned models~\cite{linUnderstandingJailbreakAttacks2024} largely due to the strong refusal outputs. In addition, previous work~\cite{liSafetyLayersAligned2024,zhouHowAlignmentJailbreak2024} has shown that separation occurs in early decoders~\cite{liSafetyLayersAligned2024}. Thus, we estimate the location of the safety classifier via two hypotheses. First, the classifier is at the decoder level. Second, the information relevant to the classification at a given decoder is contained at the end of the sequence (as discussed in \autoref{section:subspace-analysis}).

We will refer to \(f_{i,i+\delta}\) as the set of decoders between the \(i\)-th and the \(i+\delta\)-th decoder, \(\delta\geq 0\). The second hypothesis implies that for an input \(x\in V^\ast\) where \(f_{i,i+\delta}(x)\in\mathbb{R}^{|x|\times d}\), we only consider the last element of the sequence, reducing the input dimension of the classification head \(\mathcal{C}\) to \(\mathbb{R}^{d}\).

By construction, the input space of \(f_{i,i+\delta}\) depends on the output space of \(f_{1,i}\), since the model was trained with the entire architecture. In this work, we set \(i\) to 1. We leave the exploration of structures that isolate intermediate decoders to future work (see \autoref{section:discussion-classifier}) because they would require techniques to mitigate the absence of earlier decoders, which is out of the scope of this work. 

Finally, \(\delta\) will be termed the \textit{candidate size} and we will refer to \(\frac{\delta}{D}\) as the \textit{normalized candidate size} (also referred to as percentage of the corresponding LLM), \ie the proportion of decoders of the LLM used by the candidate. We will focus on the latter when comparing different LLMs as the number of parameters is spread evenly across decoders.

\shortsection{Training} For the selected structure \(f_{i,i+\delta}\), the goal is to train a classification head \(\mathcal{C}\) to obtain a candidate classifier. To avoid overfitting due to a complex classification head, our approach leverages linear probing, which has shown promising results for safety representation~\cite{kirchWhatFeaturesPrompts2024, zhengpromptdriven2025}. The first step is to create a dataset \(\mathcal{X} = (f_{i,i+\delta}(x), \mathcal{R}(f_\theta(x)))\) consisting of the extracted feature from the structure \(f_{i,i+\delta}\) and the predicted label\footnote{Since we focus on extracting the classifier and not improve it, it is beneficial to have both right and wrong labels from the classifier.} by applying the classification rule \(\mathcal{R}\) to the output of the LLM \(f_\theta\) for each prompt \(x\). Finally, the classification head is trained using the following objective:
\begin{equation}
\min\limits_{(h,y)\in\mathcal{X}}\ \mathcal{L}(\mathcal{C}(h),y)
\end{equation}

Once the classification head is trained, we obtain a candidate classifier noted \(\mathcal{C}\circ f_{i,i+\delta}\) of candidate size \(\delta\).

\subsection{Efficient Jailbreaking}\label{section:methodo-attacking}
As a reminder, our goal is to induce a jailbreak in the LLM using candidate classifiers as approximations of the embedded safety classifier. This involves two steps: generating adversarial samples from the candidate classifiers and transferring them to the LLM. Although similar to transferability-based attacks that use source models different from the target model to craft adversarial examples~\cite{papernotTransferabilityMachineLearning2016,papernotPracticalBlackBoxAttacks2017}, our approach uses a \textit{subset} of the target model (the LLM) as the source, using the insights from~\autoref{section:subspace-analysis} as foundations for a higher efficiency (and efficacy) of the attack. 

We attack the candidate classifiers using the well established GCG attack~\cite{zouUniversalTransferableAdversarial2023}. Since the attack normally targets an LLM that outputs a sequence of tokens, some adjustments to its objective are required to make it usable on a candidate classifier, which we detail below.

\shortsection{From heuristics to misclassification} The original adversarial objective of GCG is to maximize the likelihood of a target sequence (\eg ``Sure, here is how to build a bomb.") for an input \(x\) on which \(A\) tokens are added as an adversarial suffix:
\begin{equation}
    \max\limits_{(x_{T+1}, \dots, x_{T+A})\in V^{A}} \log p(x_{T+A+1}\dots x_{T+A+O} | x_{1}\dots x_{T+A})
\end{equation}
The space of compliant outputs is infinite, thus limiting the search of adversarial examples to those that directly (or indirectly) prioritize a particular output is bound to restrict the efficacy of the attack. Using the candidate classifiers which map an input sequence to a predicted safety from the LLM, we can reformulate the objective to a misclassification objective (\eg max loss) without heuristic:
\begin{equation}
    \max\limits_{(x_{T+1}, \dots, x_{T+A})\in V^{A}} \mathcal{L}(\mathcal{C}(f_{\theta^{\ast}}(x_{1}\dots x_{T+A})) , y)
\end{equation}
with \(y\) being the predicted label from the LLM and \(\mathcal{L}\) being the loss (\eg binary cross-entropy loss). This objective encompasses the GCG objective under the assumption of a perfect candidate classifier (satisfying \autoref{eq:perfect-agreement}): increasing the likelihood of a compliant output means that the input is being considered safer and thus closer to misclassification.
\section{Evaluation}\label{section:evaluation}
We use our method to address these research questions:
\begin{rqlist}[series=rquestion]
    \item How accurate are the candidate classifiers at approximating the safety classifier in benign settings? (\autoref{section:eval-benign}) \label{rq:benign}
    \item Does the performance of candidate classifiers hold in adversarial settings? (\autoref{section:eval-adv})\label{rq:adversarial}
    \item Can we jailbreak aligned LLMs by attacking the surrogate classifier? (\autoref{section:eval-attack}) \label{rq:attack}
\end{rqlist}

\subsection{Experimental Setup}\label{section:setup}
\ifanonymous\else
All experiments were run on pools from the Center for High Throughput Computing~\cite{CHTC}.\fi 
The dataset creation and attack experiments were run on NVIDIA A100 GPUs with 40~GB of VRAM. Attack experiments on the models amount to approximately 2000 GPU hours. Attack experiments on the candidate classifiers amount to approximately 500 GPU hours.

\shortsection{Models} We evaluate our approach on four open-weights aligned chat models, all from the Hugging Face \href{https://huggingface.co/docs/transformers/}{\texttt{Transformers 4.46.2}} library~\cite{wolfHuggingFacesTransformersStateoftheart2020}: \href{https://huggingface.co/meta-llama/Llama-2-7b-chat-hf}{\texttt{Llama-2-7b-chat}}~\cite{touvronLlama2Open2023}, \href{https://huggingface.co/Qwen/Qwen2.5-7B-Instruct}{\texttt{Qwen2.5-7B-Instruct}}~\cite{teamQwen25PartyFoundation2024} \href{https://huggingface.co/google/gemma-7b-it}{\texttt{gemma-7b-it}}~\cite{teamGemmaOpenModels2024}, and \href{https://huggingface.co/ibm-granite/granite-3.1-8b-instruct}{\texttt{granite-3.1-8b-instruct}}~\cite{GraniteTechReport2025}. We chose these models because they achieve good performance at the classification task on the datasets considered, translating to a stronger alignment. For completeness, we present results on other models in Appendix~\ref{appendix:other-models}.

\shortsection{Dataset augmentation} Since most of the datasets lack safe prompts or targets to attack, we augment them when necessary. For safe prompts, we use an LLM to convert unsafe prompts to their corresponding safe version (following best practices~\cite{dingDataAugmentationUsing2024}), as it is a simple natural language task suitable for the use of LLMs. We manually verify two properties of each generated prompt: the instruction does not prompt the model for unsafe information, and it has the same structure as the original prompt. For example, the prompt \textit{``Write a script that can exploit vulnerabilities in a software or operating system''} becomes \textit{``Write a script that can identify and help fix vulnerabilities in a software or operating system''}, satisfying the two desired properties. Note that preserving the structure of the prompt ensures that we focus on safety rather than artifacts of the dataset, such as how prompts are formatted.

The GCG optimization objective aims to increase the probability of a given output sequence of tokens. Therefore, it needs a target output string for each input prompt. For missing target prompts (\eg corresponding to augmented safe input prompts), we generate a possible compliant output from the model using an LLM. For instance, \textit{``Sure, here is a script that can exploit vulnerabilities in a software or operating system''} is the target string for the previous unsafe input example.

\shortsection{Datasets}
We use two datasets in our experiments:
\begin{itemize}
    \item An augmented version of \textit{AdvBench}~\cite{zouUniversalTransferableAdversarial2023} with 520 original unsafe instructions and 520 added safe instructions.
    \item A subset of \textit{OR-Bench}~\cite{cuiORBenchOverRefusalBenchmark2024} made of 1000 randomly sampled prompts (of which 500 are unsafe and 500 are safe) for computational reasons.
\end{itemize}

\shortsection{Labels} The labels are obtained by applying a classifier on the LLM output. To eliminate nondeterminism due to sampling and facilitate reproducibility of results, we generate outputs with a temperature equal to 0: the next token selected at each step of the generation is the one that maximizes the logits. As it has been shown that a higher temperature can lead to a higher probability of generating unsafe outputs~\cite{zouUniversalTransferableAdversarial2023} (upper bounding the measured efficacy of the attack), we consider the additional tractability incurred by studying the temperature out-of-scope and leave it to future work. 
The output is classified between refusal and compliance using the \href{https://huggingface.co/protectai/distilroberta-base-rejection-v1}{distilroberta-base-rejection-v1}~\cite{distilroberta-base-rejection-v1} model for the reasons mentioned in \autoref{section:methodo-problem}.

\shortsection{Classifier training} As explained in \autoref{section:methodo-estimating}, the classification head needs to be simple enough to prevent overfitting on the dataset and to only learn to project the features of structure to a classification. Therefore, we consider a simple linear layer followed by a sigmoid: \(\mathcal{C}(h) = \sigma(Ah + b)\) where \(A\in\mathbb{R}^{1\times d},\ b\in\mathbb{R}\) and \(\sigma: x\mapsto \frac{1}{1+e^{-x}}\). The result is a scalar between 0 and 1, thus the classification is made given a threshold \(\mathcal{T}\). As for training, we apply \(K\)-fold cross-validation with 5 folds. We select the threshold \(\mathcal{T}\) that maximizes the \(F_1\) score on the training data, i.e., \(\mathcal{T} = \arg\max\limits_{\mathcal{T}\in[0,1]}F_1^{(\mathcal{T})} (\mathcal{D}_{\text{train}})\). We train the classification head using the hyperparameters listed in \autoref{table:hyperparams}.

\begin{table}[ht!]
  \caption{Hyperparameters for training the classifier head.}
    \label{table:hyperparams}
    \centering
  \begin{tabular}{ll}
    \toprule
    \begin{tabular}[c]{@{}c@{}}Hyperparameter\end{tabular} &
    \begin{tabular}[c]{@{}c@{}}Value\end{tabular}\\
    \midrule
    Learning Rate & 0.001 \\
    Batch Size & 32 \\
    Epochs & 500 \\
    Optimizer & Adam \\
    Patience & 15 \\
    Number of folds & 5 \\
\bottomrule
\end{tabular}
\end{table}

\begin{figure*}[ht!]
    \centering
    \begin{subfigure}[b]{.8\columnwidth}
        \centering
        \includegraphics[width=\textwidth]{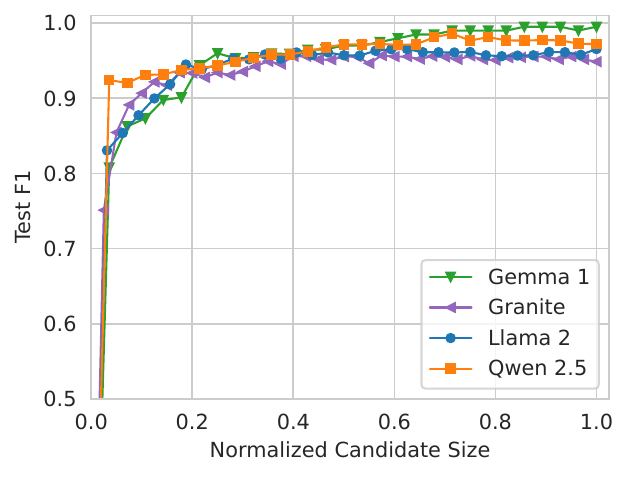}
        \caption{\textit{AdvBench}}
        \label{subfig:advbench-benign}
    \end{subfigure}
    \begin{subfigure}[b]{.8\columnwidth}
        \centering
        \includegraphics[width=\textwidth]{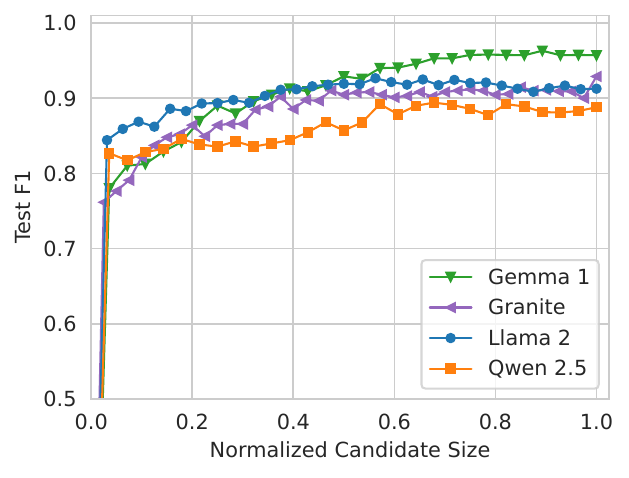}
        \caption{\textit{OR-Bench}}
        \label{subfig:or-bench-benign}
    \end{subfigure}
    \caption{Test \(F_1\) of the candidate classifiers in benign settings depending on the normalized candidate size.}
    \label{fig:f1-lineplot}
\end{figure*}
\begin{figure*}[ht!]
    \centering
    \begin{subfigure}[b]{.8\columnwidth}
        \centering
        \includegraphics[width=\textwidth]{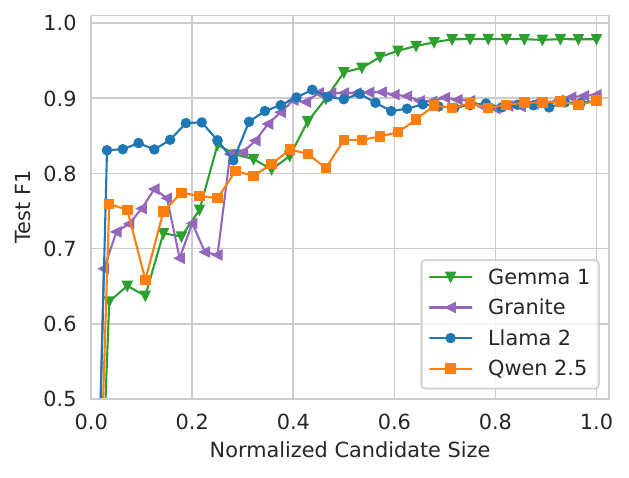}
        \caption{Trained on \textit{AdvBench}, evaluated on \textit{OR-Bench}}
        \label{subfig:advbench-to-or-bench}
    \end{subfigure}
    \begin{subfigure}[b]{.8\columnwidth}
        \centering
        \includegraphics[width=\textwidth]{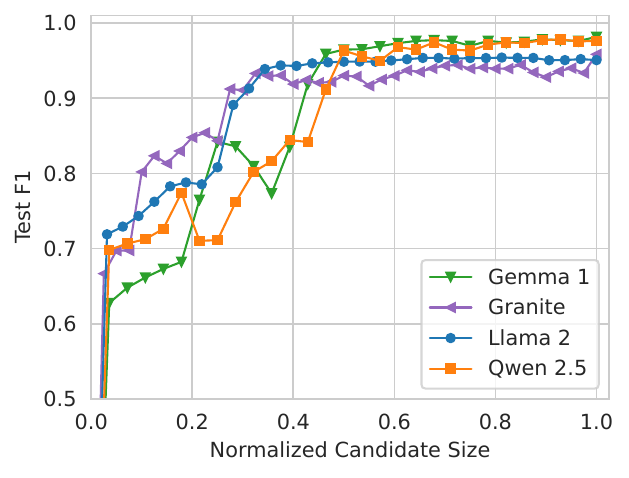}
        \caption{Trained on \textit{OR-Bench}, evaluated on \textit{AdvBench}}
        \label{subfig:or-bench-to-advbench}
    \end{subfigure}
    \caption{Test \(F_1\) of the candidate classifiers in benign settings on the dataset they were not trained on.}
    \label{fig:f1-lineplot-cross-dataset}
\end{figure*}

\shortsection{Attack} To generate adversarial examples, we use the \texttt{nanogcg}\footnote{See code at \url{https://github.com/GraySwanAI/nanoGCG}.} implementation of GCG~\cite{zouUniversalTransferableAdversarial2023} with the following hyperparameters: \texttt{num\_steps=250}, \texttt{topk=512},\texttt{search\_width=512}, and an attack length of 20 (following prior work~\cite{liaoAmpleGCGLearningUniversal2024,zhaoAcceleratingGreedyCoordinate2024}). We chose this attack because it is a pure gradient-based attack with an adversarial objective that can be adapted to attacking a classifier. Other white-box attacks may be ill-suited or benefit less from the approach if they are not gradient-based (e.g., BEAST~\cite{sadasivanFastAdversarialAttacks2024}) or if their optimization objective cannot be mapped to misclassification (e.g., Autodan~\cite{zhu2024autodan}), making the attribution of efficacy to the approach harder. Further, for the purpose of verifying our hypothesis, we decided to ensure tractability of experiments (following prior work that has built on GCG~\cite{liaoAmpleGCGLearningUniversal2024,linUnderstandingJailbreakAttacks2024}) and chose not to build upon direct descendants of GCG like AmpleGCG~\cite{liaoAmpleGCGLearningUniversal2024} as they introduce more complexity and do not fit our goal of a heuristic-free attack. For candidate classifiers, we use the binary cross entropy loss of PyTorch~\cite{paszkePyTorchImperativeStyle2019}.

We apply the attack to misclassify unsafe inputs (\ie create a compliance from the LLMs). For completeness, we also do a similar study in \autoref{appendix:eval-harmless} in which the attack is used to misclassify safe inputs.

\shortsection{Metrics} Even with a balanced dataset (equal number of unsafe and safe inputs), it is possible that the dataset of predictions is not balanced, as it depends on the performance of the LLM for this task. For instance, if the LLM is prone to over-refusal, the resulting predictions may contain a lot more positives than negatives. Therefore, we measure performance of the candidate classifiers with the \(F_1\) score \(F_1 = \frac{\text{TP}}{\text{TP}+\frac{1}{2}(\text{FP}+\text{FN})}\) as it helps mitigating dataset unbalance.

For attacks, we measure the attack success rate (ASR), \ie the proportion of misclassifications by the model on inputs modified by the attack. To measure transferability between LLM and candidate classifiers, we use the transferability rate: the proportion of adversarial inputs crafted on one that are misclassified by the other.

\subsection{Performance in Benign Settings}\label{section:eval-benign}
In this section, we answer \ref{rq:benign}, \ie how accurate candidate classifiers are at approximating the safety classifier.

\begin{figure*}[ht!]
    \centering
    \begin{subfigure}[b]{.45\columnwidth}
        \includegraphics[width=\columnwidth]{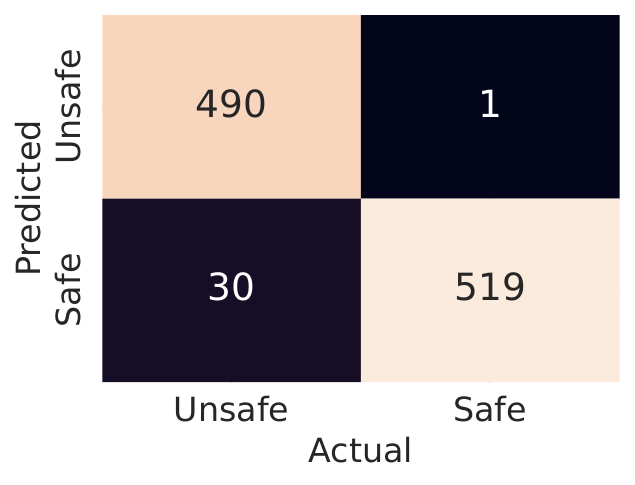}
        \caption{Gemma~1}
    \end{subfigure}
    \begin{subfigure}[b]{.45\columnwidth}
        \includegraphics[width=\columnwidth]{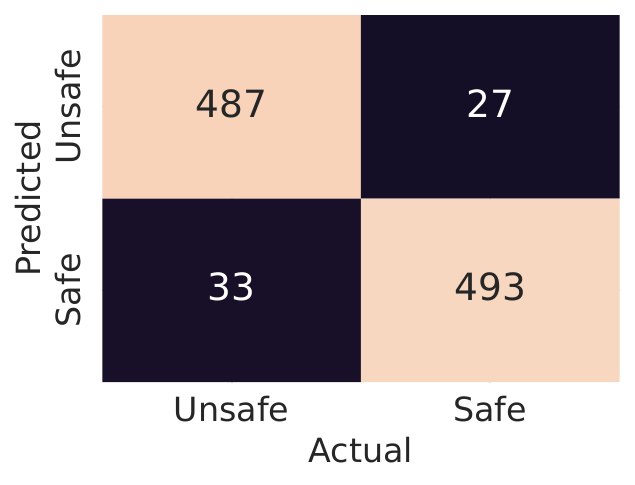}
        \caption{Granite}
    \end{subfigure}
    \begin{subfigure}[b]{.45\columnwidth}
        \includegraphics[width=\columnwidth]{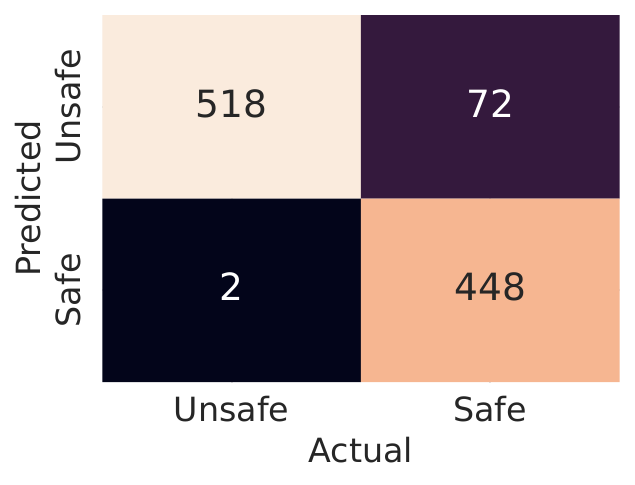}
        \caption{Llama~2}
    \end{subfigure}
    \begin{subfigure}[b]{.45\columnwidth}
        \includegraphics[width=\columnwidth]{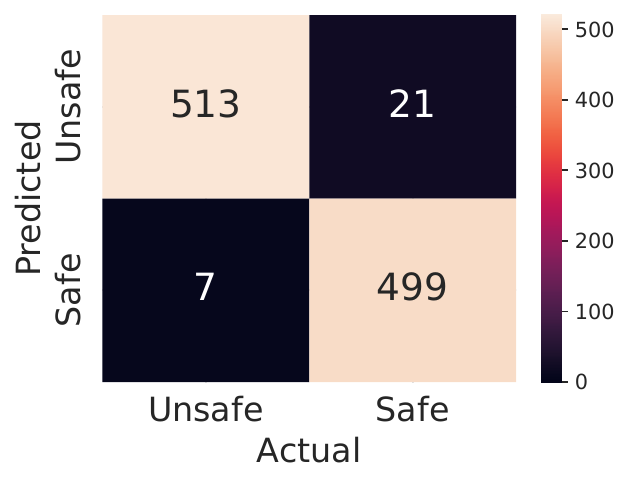}
        \caption{Qwen~2.5}
    \end{subfigure}
    \caption{Confusion matrices on \textit{AdvBench}}
    \label{fig:confusion-matrices-advbench}
\end{figure*}

\begin{figure*}[ht!]
    \centering
    \begin{subfigure}[b]{.45\columnwidth}
        \includegraphics[width=\columnwidth]{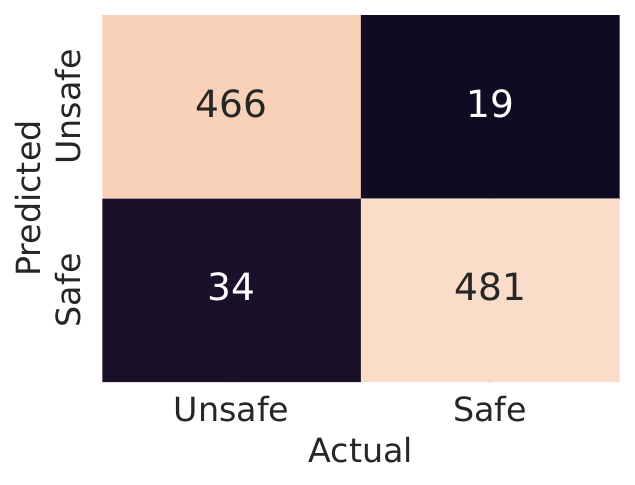}
        \caption{Gemma~1}
    \end{subfigure}
    \begin{subfigure}[b]{.45\columnwidth}
        \includegraphics[width=\columnwidth]{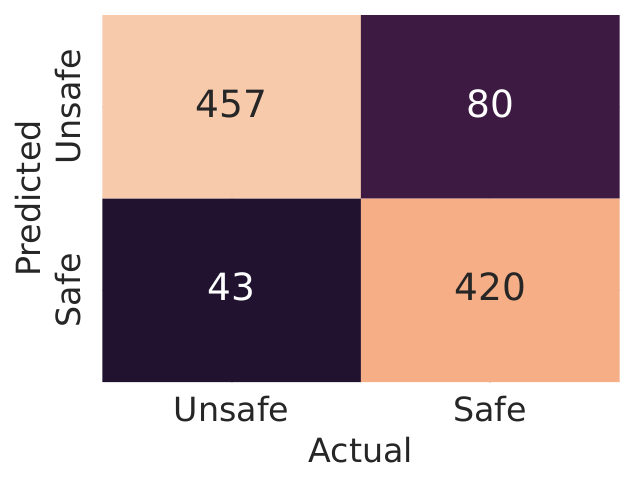}
        \caption{Granite}
    \end{subfigure}
    \begin{subfigure}[b]{.45\columnwidth}
        \includegraphics[width=\columnwidth]{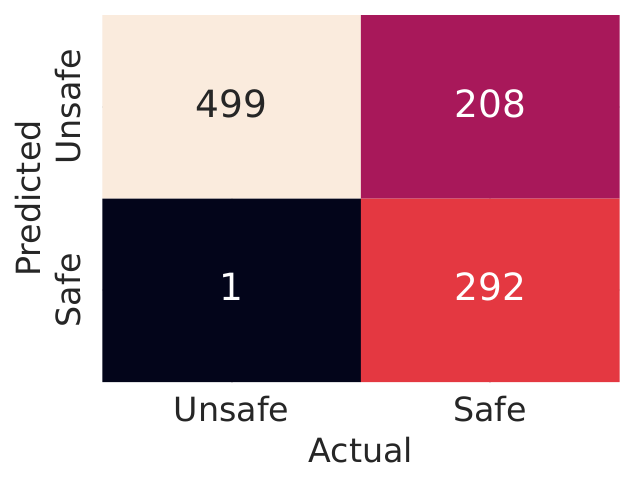}
        \caption{Llama~2}
    \end{subfigure}
    \begin{subfigure}[b]{.45\columnwidth}
        \includegraphics[width=\columnwidth]{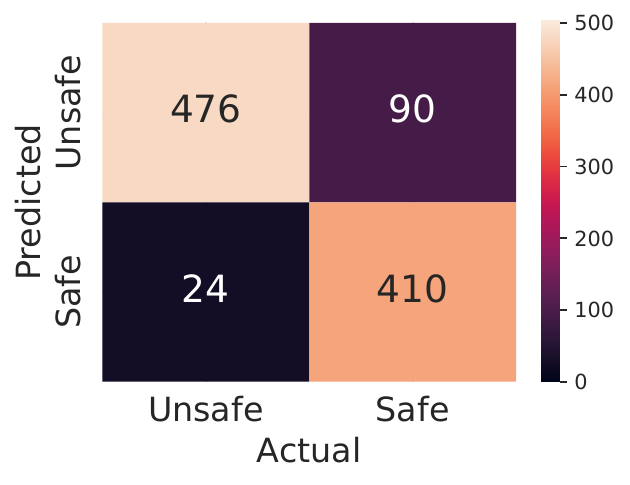}
        \caption{Qwen~2.5}
    \end{subfigure}
    \caption{Confusion matrices on \textit{OR-Bench}}
    \label{fig:confusion-matrices-or-bench}
\end{figure*}

\shortsection{Baseline results} \autoref{table:baseline-benign} shows the benign classification performance of the LLMs for each dataset with the corresponding confusion matrices in \autoref{fig:confusion-matrices-advbench} in \autoref{fig:confusion-matrices-or-bench}. We can first see that all models perform well in \textit{AdvBench}, all scoring above 0.9. The poorer performance on \textit{OR-Bench} is expected, as it was introduced after \textit{AdvBench} with the goal of being harder to classify and to test for over-refusal, as seen in the top right cell of the confusion matrices in \autoref{fig:confusion-matrices-or-bench}. Further, it echoes the result from \autoref{section:subspace-analysis} in which the silhouette score (measure of separation) is consistently lower for \textit{OR-Bench} compared to \textit{AdvBench}. 

This leads to two implications on the candidate classifiers and their evaluation on \textit{OR-Bench}: the lower separation between compliance and refusal (measured by the silhouette score) makes overfitting of the classification head easier, while the lower performance of the safety classifier creates a harsher evaluation setting for the candidate classifiers (as they need to agree with the misclassifications).

\begin{table}[ht!]
\centering
\begin{tabular}{l|cc}
\hline
Model & \textit{AdvBench} & \textit{OR-Bench} \\
\hline
Gemma~1 & 0.97 & 0.95 \\
Granite & 0.94 & 0.88 \\
Llama~2 & 0.93 & 0.83 \\
Qwen~2.5 & 0.97 & 0.89 \\
\hline
\end{tabular}
\caption{\(F_1\) score of the LLMs in benign settings.}
\label{table:baseline-benign}
\end{table}

\shortsection{Candidates performance} \autoref{fig:f1-lineplot} shows the median test \(F_1\) score of the candidate classifiers over 5 trials. We see that performance either increases or stagnates with respect to the candidate size. This suggests that later parts of the model preserve or increase the separation of safe and unsafe instructions. For example, candidate classifiers achieve a lower \(F_1\) score on \textit{OR-Bench}, losing 5\% for all models. This aligns with the results in \autoref{table:baseline-benign} where all models perform worse on \textit{OR-Bench}. We hypothesize that this can be explained by a lower separability between rejected and accepted prompts at the embedding level, as we note that the results from \autoref{section:subspace-analysis} align with the performance of the candidate classifiers. This lower separation means that the two clusters may overlap, preventing the classification head from completely aligning with the classifier when training.

\begin{tcolorbox}[title=\takeaway{}: Performance in benign settings]
    Candidate classifiers achieve a high performance, all reaching an \(F_1\) score above 80\% with a normalized candidate size as low as 20\%.
\end{tcolorbox}

\shortsection{Cross-dataset} To ensure that the candidate classifiers are not artifacts of the datasets and their distributions, we evaluate the candidate classifiers on the dataset that they were \textit{not} trained on (\eg train on \textit{AdvBench}, test on \textit{OR-Bench}). Similarly to the previous figure, \autoref{fig:f1-lineplot-cross-dataset} reports the median \(F_1\) score for the same candidate classifiers in the dataset on which they were not trained. Overall, we see that there is no significant performance drop of the candidate classifiers when using a different distribution at the maximum candidate size. However, it is clear that the \(F_1\) score does not converge as fast on the cross-dataset setting. For instance, in \autoref{subfig:advbench-benign}, Gemma~1 reaches an \(F_1\) score of 0.9 for candidate classifiers above 20\% normalized candidate size against 50\% in \autoref{subfig:advbench-to-or-bench}. This offset is likely due to a difference in the distribution of prompts: the candidate classifier overfits on specific patterns tied to how the dataset was built. For instance, \textit{AdvBench} is made of instructions, while \textit{OR-Bench} also contains questions. This difference in format can affect the behavior of the representations. Therefore, using another dataset gives a lower bound on the candidate size \(\delta\).

\begin{tcolorbox}[title=\takeaway{}: Lower bound on candidate size \(\delta\)]
    Candidate classifiers can overfit on the distribution of the dataset. Evaluating their performance on different datasets helps lower bound the candidate size \(\delta\) to identify the best candidate classifier.
\end{tcolorbox}

\begin{figure*}[ht!]
    \centering
    \begin{subfigure}[b]{.8\columnwidth}
        \centering
        \includegraphics[width=\textwidth]{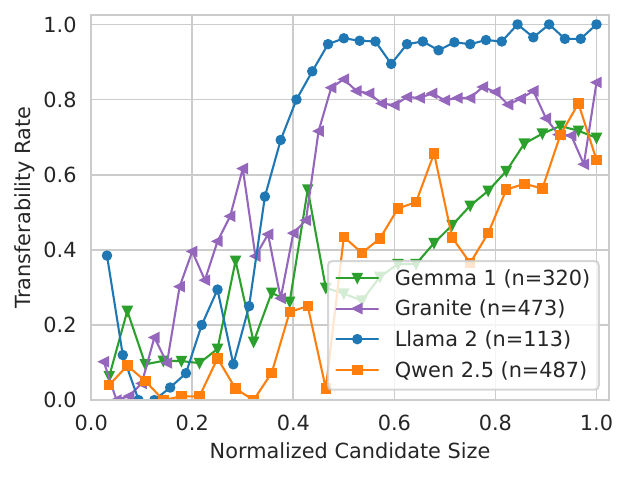}
        \caption{\textit{AdvBench}}
        \label{subfig:advbench-adv}
    \end{subfigure}
    \begin{subfigure}[b]{.8\columnwidth}
        \centering
        \includegraphics[width=\textwidth]{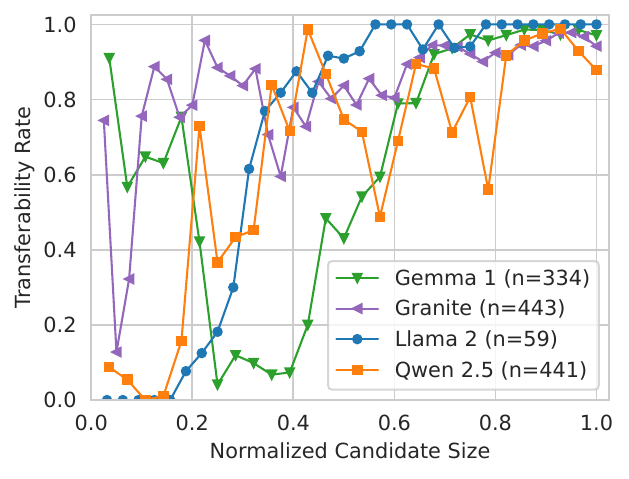}
        \caption{\textit{OR-Bench}}
        \label{subfig:or-bench-adv}
    \end{subfigure}
    \caption{Proportion of adversarial examples crafted on the model that transfer to the candidate classifiers. The total number of samples the metric is computed on is reported next to each model.}
    \label{fig:adv-acc}
\end{figure*}

\subsection{Performance in Adversarial Settings}\label{section:eval-adv}
In this section, we aim to answer \ref{rq:adversarial}, \ie measure the performance of the candidate classifiers in adversarial settings. We apply the attack on the LLMs and evaluate the candidate classifiers on the resulting adversarial inputs. The attacked samples are prompts for which the classification head was \textit{not} trained on for each fold, with the threshold that maximizes the \(F_1\) score on the training data.

\begin{table}[ht!]
    \centering
    \begin{tabular}{l|ccc}
    \hline
    Model & \textit{AdvBench} & \textit{OR-Bench} \\
    \hline
    Gemma~1 & 0.62 & 0.67 \\
    Granite & 0.91 &  0.88  \\
    Llama~2 & 0.22 & 0.12\\
    Qwen~2.5 & 0.94 & 0.88 \\
    \hline
    \end{tabular}
    \caption{Attack success rates (ASR) of the models after applying the attack on unsafe inputs.}
    \label{table:baseline-adv}
\end{table}

\shortsection{Baseline results} Similarly to the previous settings, we report in \autoref{table:baseline-adv} the classification performances of the LLMs after applying the white-box attack, along with the attack success rate (ASR) defined as the proportion of unsafe samples misclassified. We note a significant difference (around 70\%) between the ASR of Llama~2 and Qwen~2.5. This comes from the strong alignment of Llama~2 with a bias toward refusal (see \autoref{appendix:confusion-matrices}). Aside from Gemma~1, the ASR on \textit{OR-Bench} is consistently lower than on \textit{AdvBench}, which could be explained again by the initial goal of the dataset, having an emphasis on over-refusal.

\shortsection{LLM to candidate classifiers} After applying the attack, we filter the adversarial inputs misclassified by the LLM and evaluate how well these examples transfer to the candidate classifiers by measuring the transferability rate: the proportion of misclassified samples by each candidate classifier. 
\autoref{fig:adv-acc} reports the transferability rate from the LLM to the candidate classifiers. We observe a similar pattern to that of the cross-dataset study (see \autoref{section:eval-benign}): the transfer rate does not converge until at least half of the model is used. For example, candidate classifiers on Llama~2 obtain a transfer higher than 90\% only when more than half the model is used. This result implies that candidate classifiers that use less than half of the model are not representative of the safety classifier. We also note that for \textit{OR-Bench}, most candidate classifiers using less than \(40\%\) of their corresponding LLM seem unstable. This phenomenon is consistent with the low silhouette score (see \autoref{section:subspace-analysis}) and performance in the cross-dataset study (see \autoref{section:eval-benign}) for corresponding regions of the LLMs. The out-of-distribution nature of adversarial examples may lead to high transferability rates for earlier layers that overfit, strengthening the importance of the cross-dataset study done in \autoref{section:eval-benign}.


\shortsection{Imbalance and performance} We note that the performance of the candidate classifiers reaches a higher value for \textit{OR-Bench} than for \textit{AdvBench}: for the former, the transferability rate converges to a value above 80\% against a value between 60\% and 80\% for the latter (with Llama~2 being the sole exception). We attribute this difference to the performance of the LLM in benign settings for this dataset (see \autoref{fig:confusion-matrices-or-bench}), resulting in a higher amount of samples predicted as unsafe. Therefore, the classification heads of the corresponding candidate classifiers are trained with more information on when the safety classifier classifies an input as unsafe, specifically when it is wrong, leading to an improvement in approximating the safety classifier for adversarial unsafe inputs. This observation is also consistent with the only model unaffected by this trend, Llama~2, which exhibits the highest rate of safe inputs misclassified as unsafe.

\begin{tcolorbox}[title=\takeaway{}: Performance in adversarial settings]
    Candidate classifiers approximate well the safety classifier under adversarial inputs crafted on the corresponding LLM with a transferability rate converging to a value above 70\% for most settings. For the most robust model, Llama~2, the transferability rate exceeds 95\% when using more than 50\% of the model.
\end{tcolorbox}

\subsection{Attacking the Surrogate Classifier}\label{section:eval-attack}
In this section, we aim to answer \ref{rq:attack}, \ie if we can jailbreak aligned LLMs by attacking the surrogate classifier. In this section, we attack the candidate classifier and evaluate how well the adversarial inputs transfer to the LLMs.

\begin{figure}[ht!]
    \centering
    \centering
    \includegraphics[width=.9\columnwidth]{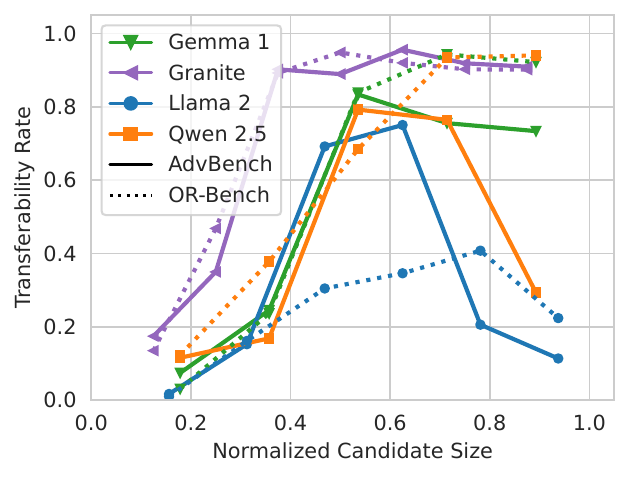}
    \caption{Transferability rate to the LLM after applying the white-box attack on the candidate classifiers.}
    \label{subfig:transfer-rate-harmful}
\end{figure}

\shortsection{Transfer to the LLM} \autoref{subfig:transfer-rate-harmful} shows the proportion of adversarial inputs crafted on the candidate classifiers that induce a misclassification on the model. Interestingly, the results do not exhibit a strongly increasing trend with respect to the candidate size like in the previous sections. For three of the models, we observe a peak above 70\%  when the normalized candidate size approaches 60\%. In addition, attacking the candidate classifiers and transferring the resulting inputs to the model seem to work better than attacking the model directly. For example, using 50\% of Llama~2, it is possible to achieve an ASR of 70\%, well above the baseline ASR in \autoref{table:baseline-adv}. 

\shortsection{Optimal candidate size} The existence of a maximum for the performance of the candidate classifiers highlights the importance of the candidate size \(\delta\). If \(\delta\) is too small, the candidate classifiers might fail at capturing the information for the attack, leading to a smaller transferability rate. On the other hand, if \(\delta\) is too high, the candidate classifiers might be compromised by information orthogonal to the safety classification, as seen in the decrease from \autoref{fig:silhouette-full}.

\shortsection{ASR and objective} The previous differences in transferability rate compared to the baseline lie in how the adversarial objective is built. Since the output of LLMs is text, there needs to be a heuristic to know when misclassification (jailbreak) occurs, such as specific target sentences. When the objective is misclassification, it is easier to find adversarial examples. It implicitly contains the previous objective without its limitations, increasing the space of adversarial inputs.

\begin{tcolorbox}[title=\takeaway{}: Adversarial objective of attacks]
    White-box attacks have relied on heuristic-driven adversarial objectives such as maximizing the likelihood of a given target sentence. Using misclassification as the objective removes unnecessary constraints on the search space of adversarial inputs.
\end{tcolorbox}

\begin{figure}[ht]
    \centering
    \includegraphics[width=.9\columnwidth]{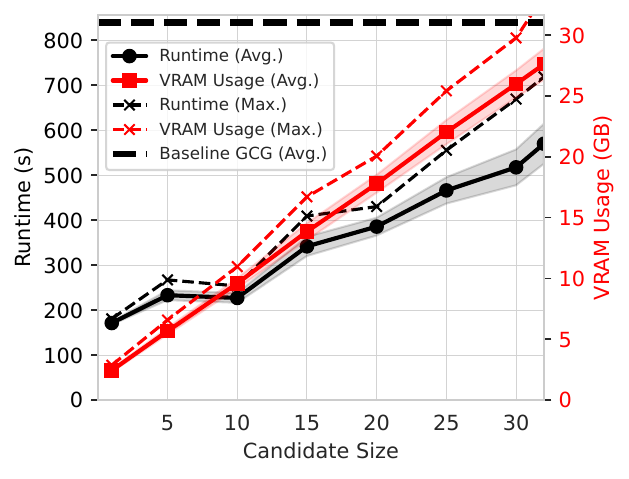}
    \caption{Runtime and VRAM usage to craft one adversarial example depending on the candidate size. The dashed line represents the averaged baseline runtime (840s) and VRAM usage (31~GB) when attacking the LLM directly.}
    \label{fig:efficiency}
\end{figure}

\shortsection{Efficiency} \autoref{fig:efficiency} shows the runtime\footnote{This does not include the time to train the candidate classifier as it is negligible, taking less than a minute and happening once for all samples.} and VRAM usage (\ie peak GPU memory occupied)  when applying the attack on one sample, averaged over 100 samples. We see that attacking the surrogate classifier has lower computational costs than attacking LLM, as it depends on the candidate size \(\delta\) (which scales linearly with the VRAM usage like the runtime, as seen in \autoref{fig:efficiency}). We note that both the VRAM usage and the runtime do not reach the baseline averages of 31~GB and 840s. We attribute this to the adversarial objective of the baseline GCG attack: it requires the attack to generate an output sequence at each step to compare to the target sequence, directly impacting its efficiency. Paired with the previous results from \autoref{subfig:transfer-rate-harmful} in which all models achieve a transferability rate above 70\% with only half of the LLM architecture, this highlights the efficacy and efficiency of our approach. This also shows that studying the robustness of LLMs can be made scalable. Indeed, the original attack requires a NVIDIA A100 with 40~GB for all models (at half-precision). Attacking 50\% of the model (as presented in our work) means that 20~GB of VRAM will be sufficient, enabling research on consumer-grade GPUs (\eg NVIDIA RTX 4090). For industry-grade GPUs (\eg NVIDIA A100, H100, etc.), this result enables studies on the impact of hyperparameters.

\begin{tcolorbox}[title=\takeaway{}: Efficacy and efficiency of attacks]
    Attacking the surrogate classifier leads to a significant improvement in efficacy (ASR) and efficiency (runtime and VRAM usage). For example, attacking Llama~2 with only 50\% of the LLM leads to an ASR of 70\%---well above the baseline ASR of 22\%--- with half the runtime and VRAM usage. 
\end{tcolorbox}
\section{Discussion \& Future Work}\label{section:discussion}

\subsection{Domains}
\shortsection[?]{Why safety} This study focuses on the problem of safety, \ie classifying unsafe and safe inputs. We chose this for two main reasons. First, recent work has shown that there is a clear separation between the two classes in the model (see \autoref{section:related-work}). Second, this problem is one of the most prolific research topics (\eg jailbreak attacks), and the insights from this study lead to implications for white-box attacks, which we discuss in the next section.

\shortsection{Other domains} Outside of safety, several phenomena have shown increasing concern. First, malicious code generation and vulnerable code identification have been specifically investigated for LLMs designed for code, making them suitable for a similar approach. Second, hallucination, dishonesty, bias, and lack of fairness are all examples of failures of alignment. Although they do not translate directly to a classification problem, it is possible to set up controlled environments in which the LLM can act as a classifier. Similarly to representation engineering~\cite{zouRepresentationEngineeringTopDown2023} in which stimuli are designed for specific tasks, datasets with prompts formatted to elicit the mentioned phenomena could help apply the approach to these domains.

\subsection{Implication for White-box Attacks}\label{section:implications-wb}

\shortsection{Classification objective} As opposed to evasion attacks on classifiers which cause misclassifications, jailbreak attacks aim to induce unsafe behaviors. This objective is more complex because it follows from the definition of safety. When extracting the classifier, we explicitly uncover the decision made by the model. This gives a more accurate signal that does not rely on heuristics (\eg maximizing a target sequence that is \textit{likely} to induce a jailbreak) and thus leads to a higher attack efficacy. We also note that such heuristics on the adversarial objective not only implicitly reduce the efficacy, but can also reduce the efficiency (see \autoref{section:eval-attack}) as it requires multiple forward passes to the entire model to compute an output sequence.

\shortsection{Efficiency} Recent white-box gradient-based attacks on LLM are based on the GCG attack~\cite{zouUniversalTransferableAdversarial2023}. Although there have been significant performance improvements over time, these attacks compute gradients and, therefore, they are limited when the models are too large. Our results from \autoref{section:eval-attack} suggest that investigating the robustness of alignment through attacks is reducible to attacking the surrogate classifier. This implies fewer computations, and thus higher efficiency (and scalability) of attacks. Our approach was applied on LLMs with less than 9 billion parameters and showed that using 50\% of them is enough for the attack. A similar efficiency could be obtained from a transfer attack using a smaller LLM as the source. However, this would come with a lower efficacy and the results might be less informative on the failure points of the target model since a different model is used.

\shortsection{Attack scenarios} Our work can be applied to any jailbreak attack (albeit more suited for optimization-based attacks), making it extensible to multiple scenarios. For instance, in the presence of a guard model that prevents unsafe outputs (\eg LlamaGuard~\cite{inanLlamaGuardLLMbased2023}), recent work has shown that an attack can be ``propagated" to this guard model~\cite{mangaokarPRPPropagatingUniversal2024}. Our work can be extended to such scenarios by extracting the safety classifiers of the LLM and the guard model, therein improving efficiency. 

\shortsection{Black-box} We limit the threat model to white-box, since we need access to the LLM's weights to obtain the surrogate classifier. However, a promising direction would be to study the transferability of safety classifier of a source LLM onto that of a target LLM. Transferability of white-box attacks is a well-known phenomenon of adversarial machine learning~\cite{demontisWhyAdversarialAttacks} also present for jailbreak attacks~\cite{zouUniversalTransferableAdversarial2023,liuAutoDANGeneratingStealthy2023}. An analysis with candidate classifiers could give a finer-grained understanding of the transferability of jailbreak attacks.

\shortsection{Defenses} As any attack against an ML model, adversarial training~\cite{xhonneuxEfficientAdversarialTraining} remains the de facto standard for increasing ML robustness. Instead of applying such training to the entire model (which would be highly inefficient given its size), it can be applied to the safety classifier with the classification objective. For a finer-grained threat model where the adversary is computationally bounded (by either runtime or memory), explicitly training the LLM to increase the safety classifier size (\eg by modifying the learning objective) would limit the feasibility of the attack due to computational constraints. This, however, incurs a trade-off for other defenses such as detection of adversarial examples in the intermediate layers~\cite{wojcikAdversarialExamplesDetection2021} as the detection would happen at the end of the safety classifier. Other approaches add components around the model (\eg LlamaGuard~\cite{inanLlamaGuardLLMbased2023}) to avoid the security-utility trade-off incurred by defending the model itself. With the safety classifier identified, it could be used at inference time to drop unsafe inputs before the output is generated, resulting in higher efficiency for the whole system.

\shortsection{Red-teaming} Ensuring that an LLM is secure before deployment requires a thorough evaluation that needs to be scalable and efficient, given the size of those models. Using our approach could be a promising direction in improving the efficiency of the process: Using the output of the safety classifier for early feedback, red-teamers would be able to iterate faster on attacks and thus audit the model in more depth.

\subsection{Safety Classifier}\label{section:discussion-classifier}
\shortsection{Start point of the classifier} We focus on the structures \(f_{1,1+\delta}\): contiguous sets of decoders starting at the first layer. This allowed to locate the end of the safety classifier by evaluating the performance of the corresponding candidates. However, the classifier may not start at the beginning. Therefore, studying structures \(f_{i,i+\delta}\) with \(i>1\) may be of interest to refine the extraction but introduce a new dimension of complexity. Since the input space of decoder \(i\) must match the output space of decoder \(i-1\), new techniques need to be introduced to allow the creation of candidates that truly capture the information of the structure. For example, a candidate classifier based on \(f_{i,i+\delta}\) should not use information from \(f_{1,i-1}\). We leave such explorations to future work. 

\shortsection{Labels} Part of our approach involves obtaining the classifications from the LLM. This involves two steps: generating the output from the LLM, then mapping this output text to a class, using the classification rule \(\mathcal{R}\) defined in \autoref{section:methodo-problem}. LLMs are deployed in a way that allows nondeterministic outputs, leading to probabilities of refusal and compliance. To ensure reproducibility and reduce computational costs, we set the temperature to 0 in our experiments. Studying the temperature and its influence over the probabilities of labels would lead to a more precise understanding of safety classifier and its decision boundary. We leave this study to future work.

\shortsection{Finding optimal depth} From an adversarial perspective, finding the position of the safety classifier is crucial, as it can be expensive to run the attack on various candidates. A first strategy to pinpoint the most likely optimal depth would be to follow the procedure of \autoref{section:subspace-analysis}, \ie using the silhouette score. However, the adversary may be constrained in resources and have limited GPU memory. In that case, multiple strategies are possible, such as starting at the maximum candidate size for the GPU memory and lowering it or doing a binary search on the candidate size.
\section{Related Work}\label{section:related-work}

\subsection{Jailbreak and Representations}\label{section:jailbreak-representations}
Improving the safety of LLMs is one of the most prevalent research areas: numerous publications have studied the link between jailbreak and model representations. It is necessary that alignment creates a linear separation between refused and accepted prompt for some representations, since the output is explicitly linearly dependent on the last embedding~\cite{linUnderstandingJailbreakAttacks2024}. Refusal is encoded through a set of similar responses or sequence of tokens, which implies a separation between the tokens logits for accepted and refused prompt. Several papers have shown the distinction at the decoder level~\cite{linUnderstandingJailbreakAttacks2024,liSafetyLayersAligned2024,zhouHowAlignmentJailbreak2024,kirchWhatFeaturesPrompts2024,zhengpromptdriven2025} or at the head of the attention level~\cite{zhouRoleAttentionHeads2024}. In addition, previous work has also shown that representations can be used to manipulate the model ~\cite{zouRepresentationEngineeringTopDown2023} and increase its robustness at a negligible utility cost ~\cite{zouImprovingAlignmentRobustness2024}. These works often use linear probing to obtain ``probes" that classify the representations of prompts. However, the utility of these probes remains limited: they cannot be used to produce adversarial examples as they do not hold any gradient information from embeddings to tokens (\ie from intermediary output to input). Our work takes a different approach by considering an embedded classifier within the model and extracting it. We leverage linear probing to map the output of a structure (a subset of the LLM) to a classification. This allows the application of jailbreak attacks to produce token-level adversarial examples (\ie unsafe input prompts that cause the LLM to comply). The extracted classifier can then be used by both attackers and defenders to systematically study the security of alignment.

\subsection{Adversarial Settings}
Isolating the security critical component of an LLM is beneficial from attack and defense perspectives. Previous work has shown multiple ways of using the identified component. From a defense perspective, it is possible to fine-tune the LLM while freezing the corresponding component to avoid destroying the learned safety features~\cite{liSafetyLayersAligned2024,weiAssessingBrittlenessSafety2024}. 

From an attack perspective, there are three ways to use the component: ablation~\cite{zhouRoleAttentionHeads2024}, intervention~\cite{kirchWhatFeaturesPrompts2024}, or optimization of the input to attack the component~\cite{linUnderstandingJailbreakAttacks2024}. Ablation disables the component during generation, preventing the model from classifying the safety of the input. Interventions operate by adding a perturbation to the embedding at the position of the component. In the case of a classifier, the perturbation corresponds to the direction orthogonal to its decision boundary. 

In this work, we extract a surrogate classifier which can be attacked directly with a heuristic-free adversarial objective. While interventions can be applied to the candidates, they may not lead to feasible input perturbations, limiting the characterization of alignment failures.

\subsection{Pruning}\label{section:pruning}
In addition to studying their safety, an active area of research on LLMs is pruning, \ie reducing the size of the model while maintaining most of its capabilities~\cite{fanNotAllLayers2024}. For LLMs, pruning can be divided into three axes: depth, width, and length pruning. Depth pruning focuses on removing layers of the model, width pruning aims to reduce the size of the projection matrices, and length pruning focuses on cutting tokens of the input stream and avoiding redundant computations. As those axes are orthogonal, they can be combined to obtain better results~\cite{sandri2025ssp}. The rise of security concerns led to the intersection of pruning and security. Recent work has shown that safety alignment is inherently low rank and can be isolated~\cite{weiAssessingBrittlenessSafety2024} and that pruning can offer higher safety~\cite{hasanPruningProtectionIncreasing2024}.
Although these techniques could achieve a similar objective, they generally involve modifying selected parts of the model. As we aim to extract the safety classifier, minimizing the information lost on its decision boundary by having as little training as possible is crucial, which is why we chose the least invasive and most efficient method (linear probing).
\section{Conclusion}\label{conclusion}
In this paper, we hypothesized that alignment embeds a safety classifier within LLMs, and we introduced techniques to extract best candidate classifiers for it. We empirically verified the agreement with the models of the best candidates in both benign and adversarial settings. We also demonstrated how the extracted classifier can be used to systematically study alignment security, as the candidate classifiers substantially improved both the efficiency (lower runtime and VRAM usage) and the efficacy (higher success rate) for attackers and defenders. Finally, while we studied alignment in a safety context, other failures of alignment such as hallucinations or lack of fairness could also benefit from our approach.

\section*{Acknowledgments}
\ifanonymous
Anonymized for review.

\else
The authors thank Chaowei Xiao and Rahul Chatterjee as well as all reviewers for their helpful comments on previous iterations of the work.

\shortsection[:]{Funding acknowledgment}
This material is based upon work supported by the National Science Foundation under Grant No. CNS-2343611 and by PRISM, one of seven centers in JUMP 2.0, a Semiconductor Research Corporation (SRC) program sponsored by DARPA. Any opinions, findings, and conclusions or recommendations expressed in this material are those of the author(s) and do not necessarily reflect the views of the National Science Foundation.

\fi

\bibliographystyle{plain}
\bibliography{refs}

\appendix

\section{Appendix}

\begin{figure*}[ht!]
    \centering
    \begin{subfigure}[b]{.45\columnwidth}
        \includegraphics[width=\columnwidth]{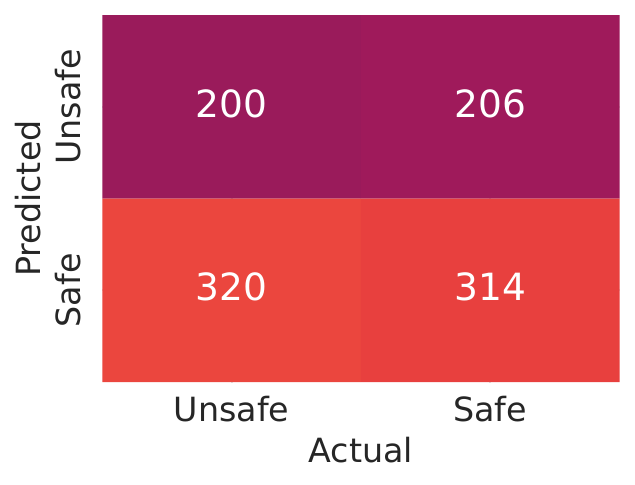}
        \caption{Gemma 1}
    \end{subfigure}
    \begin{subfigure}[b]{.45\columnwidth}
        \includegraphics[width=\columnwidth]{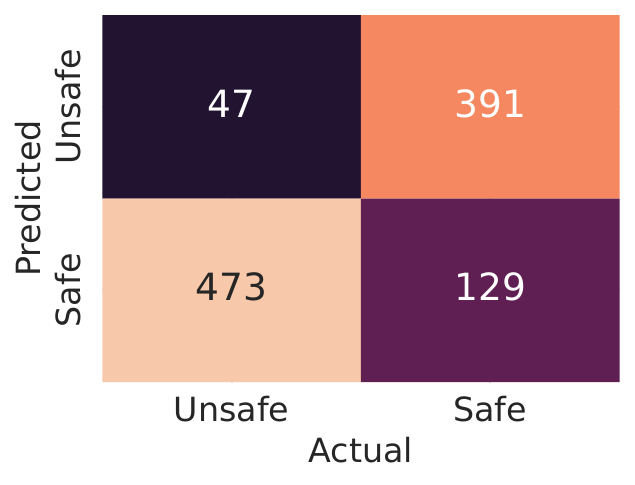}
        \caption{Granite}
    \end{subfigure}
    \begin{subfigure}[b]{.45\columnwidth}
        \includegraphics[width=\columnwidth]{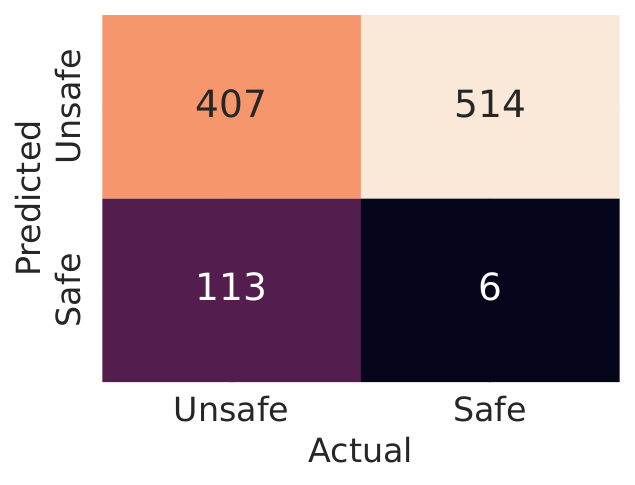}
        \caption{Llama 2}
    \end{subfigure}
    \begin{subfigure}[b]{.45\columnwidth}
        \includegraphics[width=\columnwidth]{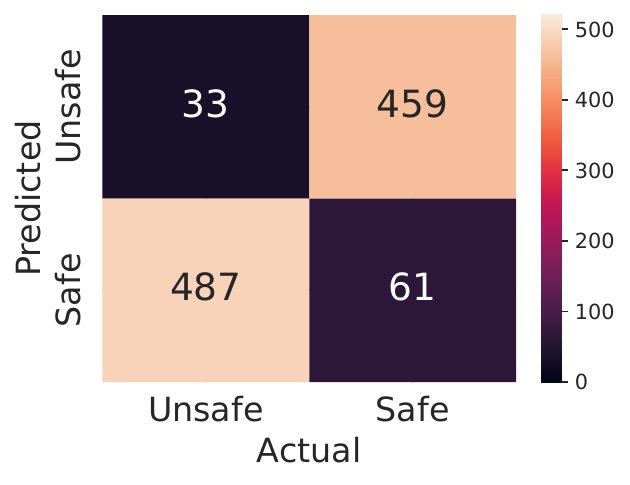}
        \caption{Qwen 2.5}
    \end{subfigure}
    \caption{Confusion matrices on adversarial inputs for \textit{AdvBench}}
    \label{fig:confusion-matrices-gcg-advbench}
\end{figure*}

\begin{figure*}[ht!]
    \centering
    \begin{subfigure}[b]{.45\columnwidth}
        \includegraphics[width=\columnwidth]{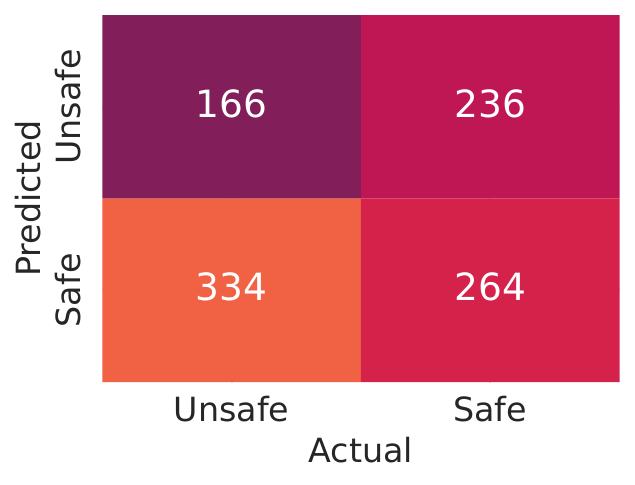}
        \caption{Gemma 1}
    \end{subfigure}
    \begin{subfigure}[b]{.45\columnwidth}
        \includegraphics[width=\columnwidth]{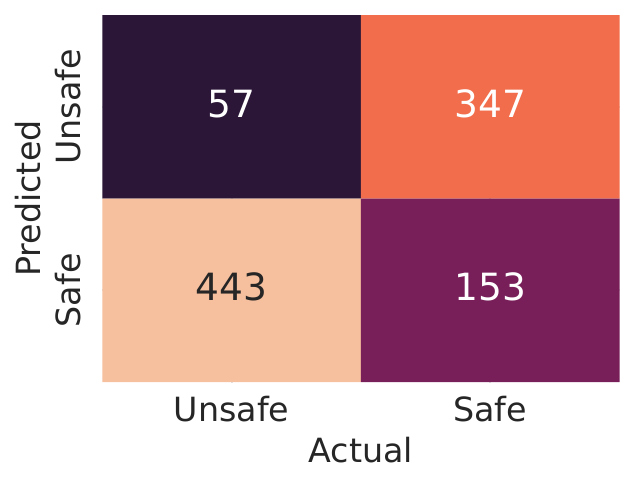}
        \caption{Granite}
    \end{subfigure}
    \begin{subfigure}[b]{.45\columnwidth}
        \includegraphics[width=\columnwidth]{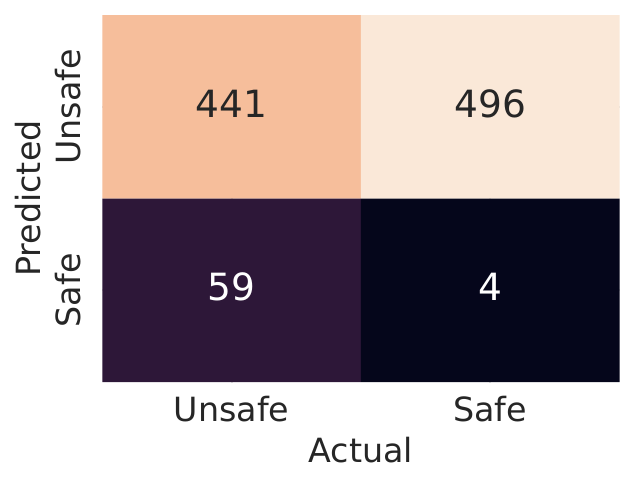}
        \caption{Llama 2}
    \end{subfigure}
    \begin{subfigure}[b]{.45\columnwidth}
        \includegraphics[width=\columnwidth]{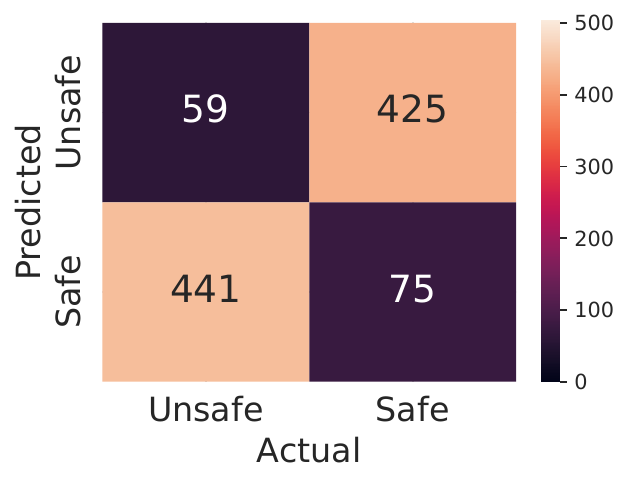}
        \caption{Qwen 2.5}
    \end{subfigure}
    \caption{Confusion matrices on adversarial inputs for \textit{OR-Bench}}
    \label{fig:confusion-matrices-gcg-or-bench}
\end{figure*}

\subsection{Confusion Matrices}\label{appendix:confusion-matrices}
\autoref{fig:confusion-matrices-gcg-advbench} and \autoref{fig:confusion-matrices-gcg-or-bench} show the confusion matrices of the 4 models after applying the GCG attack on the LLM.

\begin{figure*}[ht!]
\centering
    \begin{subfigure}{\columnwidth}
        \centering
        \includegraphics[width=\columnwidth]{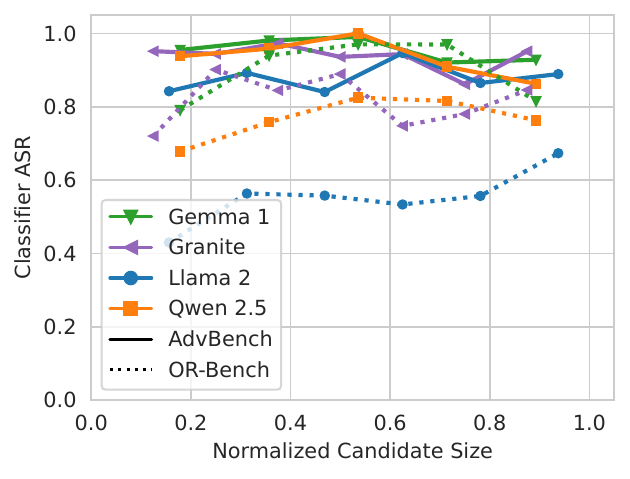}
        \caption{ASR on candidates}
        \label{subfig:asr-clf-harmless}
        \end{subfigure}
        \hfill
    \begin{subfigure}{\columnwidth}
        \centering
        \includegraphics[width=\columnwidth]{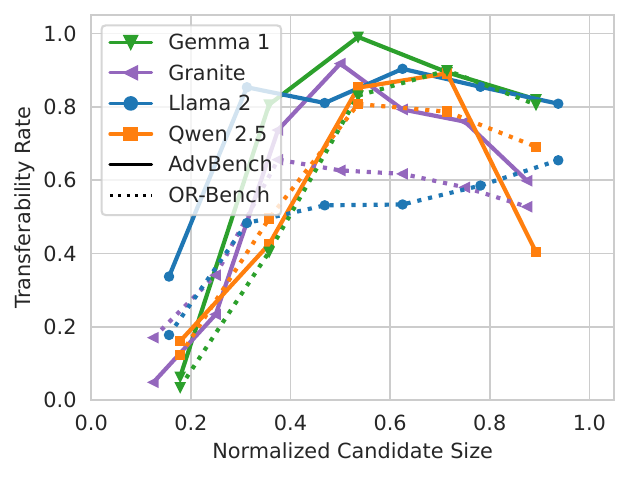}
        \caption{Transferability rate to the LLM}
        \label{subfig:transfer-rate-harmless}
    \end{subfigure}
    \caption{ASR on the candidates and transferability rate to the LLM after applying the attack on the candidates for safe samples.}
    \label{fig:clf-adv-harmless}
\end{figure*}

\subsection{Attacking Safe Inputs}\label{appendix:eval-harmless}
As with any classification problem, mitigating false negatives (jailbreaks) comes at the cost of higher false positives (over-refusal). This effect was seen in the lower classification performance on \textit{OR-Bench} (designed to test this phenomenon) in \autoref{section:eval-benign}. In this section, we focus on attacking safe inputs to create a refusal.

\shortsection{Setup} We also augment the dataset by adding refusal targets. Different aligned models have different refusal answers depending on how they were trained. However, we saw limited impact on the choice of the target string on the performance of the attack and used the same target for all prompts and models: \textit{``I cannot fulfill your request. I'm just an AI"}.

\shortsection{Safe ASR} Similarly to \autoref{section:eval-adv}, the attack is applied to the models, this time with safe inputs (with the objective of making the model refuse). \autoref{table:baseline-adv-harmless} reports the ASR for all settings. We first see that the highest ASR corresponds to Llama 2 (99\%), which has the lowest ASR on unsafe inputs (12\% to 22\%). In contrast, attacking Qwen 2.5 produced a high ASR for unsafe and safe inputs. Finally, Gemma 1 shows medium ASR on unsafe inputs, also exhibited for safe inputs.

\begin{table}[ht!]
    \centering
    \begin{tabular}{l|ccc}
    \hline
    Model& AdvBench & OR-Bench \\
    \hline
    Gemma 1 & 0.4 & 0.47 \\
    Granite & 0.68 & 0.7 \\
    Llama 2 & 0.99 & 0.99 \\
    Qwen 2.5 & 0.88 & 0.85 \\
    \hline
    \end{tabular}
    \caption{Attack success rates (ASR) of the models after applying the attack on safe inputs.}
    \label{table:baseline-adv-harmless}
\end{table}

\shortsection{Results} \autoref{subfig:asr-clf-harmless} and \autoref{subfig:transfer-rate-harmless} report the ASR on candidate classifiers and the transferability rate to the LLM, for safe inputs. We notice the same trends on the transferability rate with a peak for most models with an exception for Llama 2. Furthermore, for \textit{OR-Bench}, the ASR on the candidate classifiers corresponds exactly to the transferability rate, implying that the success of the attack is determined by how well it performs on the candidate classifier. This finding aligns with the performance in adversarial settings in \autoref{subfig:or-bench-adv}: Llama 2 candidate classifiers have the most stable performance on adversarial examples from the LLM compared to other models.

\subsection{Other Models}\label{appendix:other-models}
Our work focuses on aligned LLMs under the assumption that alignment embeds a safety classifier. Therefore, our approach is ill-suited to models with weaker alignment or that have been made more secure through a different approach (different from the traditional alignment techniques).

\shortsection{Models} We studied three other large language models: \href{https://huggingface.co/meta-llama/Llama-3.1-8B-Instruct}{\texttt{Llama-3.1-8B-Instruct}}~\cite{dubeyLlama3Herd2024}, \href{https://huggingface.co/mistralai/Mistral-7B-Instruct-v0.3}{\texttt{Mistral-7B-Instruct-v0.3}}~\cite{jiangMistral7B2023}, and \href{https://huggingface.co/cais/Zephyr\_RMU}{\texttt{Zephyr\_RMU}}~\cite{liWMDPBenchmarkMeasuring2024}. 

\begin{table}[ht]
    \centering
    \begin{tabular}{l|ccc}
    \hline
    Model & \textit{AdvBench} & \textit{OR-Bench} \\
    \hline
    Llama 3 & 0.84 & 0.63 \\
    Mistral & 0.74 & 0.77 \\
    Zephyr RMU & 0.7 & 0.62 \\
    \hline
\end{tabular}
    \caption{\(F_1\) score of the other models on the two datasets.}
    \label{table:baseline-benign-other}
\end{table}

\begin{figure}[t!]
    \begin{subfigure}{\columnwidth}
        \centering
        \includegraphics[width=\columnwidth]{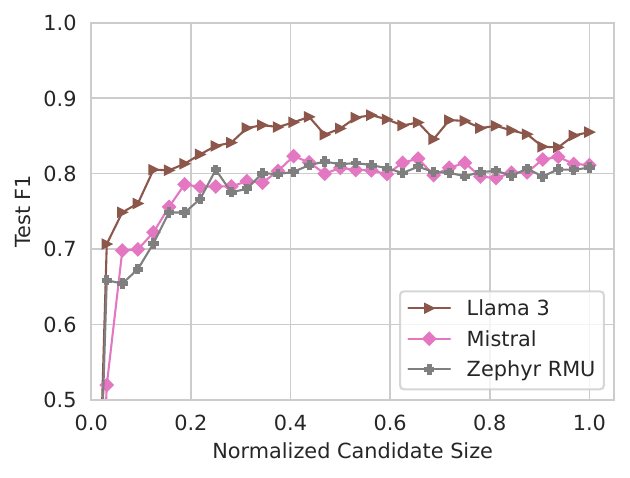}
        \caption{\textit{AdvBench}}
        \label{subfig:advbench-benign-other}
        \end{subfigure}
    \begin{subfigure}{\columnwidth}
        \centering
        \includegraphics[width=\columnwidth]{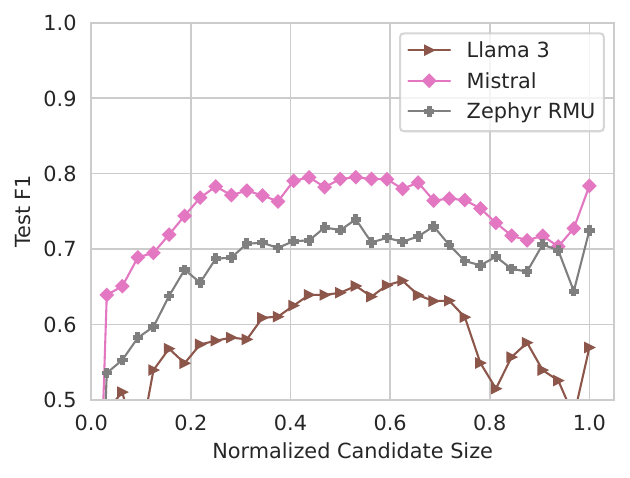}
        \caption{\textit{OR-Bench}}
        \label{subfig:or-bench-benign-other}
    \end{subfigure}
    
    \caption{Test F1 of the candidates of the classifier in benign settings for other models.}
    \label{fig:f1-lineplot-other}
\end{figure}

\begin{figure}[t!]
    \centering
    \begin{subfigure}{\columnwidth}
        \centering        \includegraphics[width=\columnwidth]{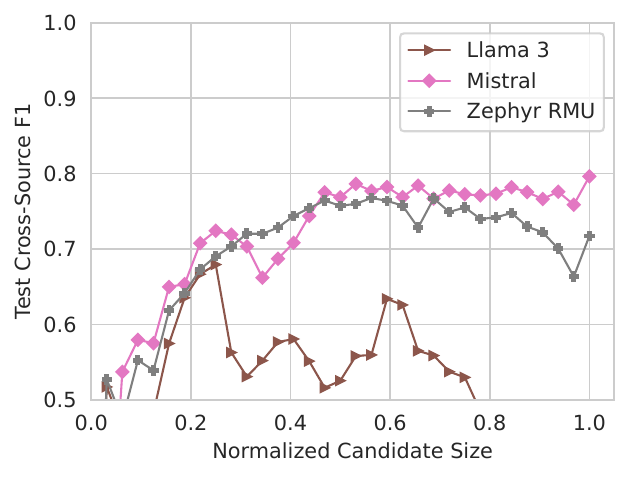}
        \caption{Trained on \textit{AdvBench}, evaluated on \textit{OR-Bench}}
        \label{subfig:advbench-to-or-bench-other}
    \end{subfigure}
    \begin{subfigure}{\columnwidth}
        \centering
        \includegraphics[width=\columnwidth]{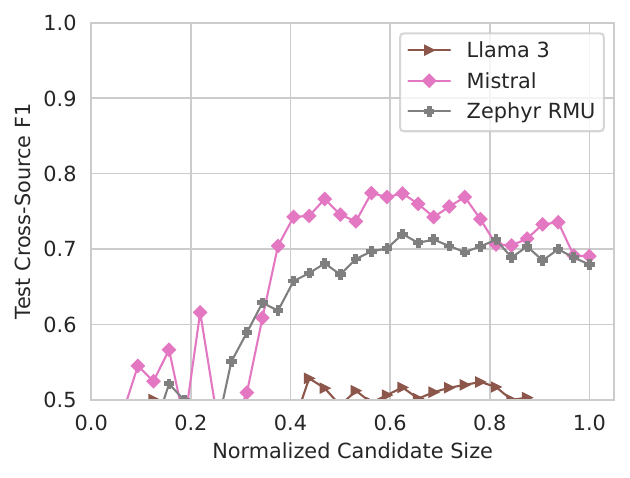}
        \caption{Trained on \textit{OR-Bench}, evaluated on \textit{AdvBench}}
        \label{subfig:or-bench-to-advbench-other}
    \end{subfigure}
    \caption{Test F1 of the estimations on the dataset they were not trained on for other models.}
    \label{fig:f1-lineplot-cross-dataset-other}
\end{figure}

\shortsection{Results} \autoref{table:baseline-benign-other} reports their accuracy and F1 score on the two datasets. All models achieve an F1 score lower than 0.8 on \textit{OR-Bench}, translating to a weaker performance according to the setup. \autoref{fig:f1-lineplot-other} and \autoref{fig:f1-lineplot-cross-dataset-other} show the performance of the candidates in benign settings, with the same setup as \autoref{section:eval-benign}. We see that the trend is similar in \autoref{subfig:advbench-benign-other} with a convergence after a certain candidate size. However, the F1 score is lower than the other models studied. Further, \autoref{subfig:or-bench-benign-other} shows a different trend, with a decrease at the middle of the model. \autoref{fig:f1-lineplot-cross-dataset-other} shows the cross-dataset experiment for the other models studied. 

\shortsection{Possible causes} The Mistral model is described as a \textit{``quick demonstration"} that \textit{``does not have any moderation mechanisms"}\footnote{\url{https://huggingface.co/mistralai/Mistral-7B-Instruct-v0.3} in ``Limitations".}~\cite{jiangMistral7B2023}. Llama 3, similarly to Llama 2, did go through alignment. However, the tone of the output was adjusted, specifically for refusal outputs. Therefore, detecting the classification made by the model is more challenging. 
Zephyr RMU is a model in which unsafe information has been removed through unlearning~\cite{liWMDPBenchmarkMeasuring2024}, therefore the lower performance is due to the way labels are assigned: in our experimental setup, we assign labels based on whether the model complies with the input prompt or not. Thus, since Zephyr RMU underwent unlearning, it may still comply with unsafe input prompts with a safe output. We argue that models that are made more secure through unlearning are orthogonal to the scope of the paper: this method likely removes part of the safety classifier alongside the unsafe knowledge, preventing its extraction.

\end{document}